\documentclass[manuscript]{emulateapj}

\usepackage{graphicx}
\usepackage{tabulary}
\usepackage{epsfig}
\usepackage{amssymb}
\usepackage{multirow}
\usepackage{appendix}
\usepackage{natbib}
\newcommand{\icarus}{Icarus\ }  
\newcommand{\beq}[1]{\begin{equation}\label{#1}}
\newcommand{\eeq}{\end{equation}}
\newcommand{\ME}{M_{\oplus}}
\newcommand{\MSUN}{M_{\odot}}
\newcommand{\Ms}{M_{\star}}
\newcommand{\Rol}{Ro^*_l}

\renewcommand{\bar}{\overline}
\newcommand{\rom}[1]{{\rm #1}}
\newcommand{\dtot}[2]{ \frac{ d #1 }{d #2} }

\newcommand{\gs}{\gtrsim}
\newcommand{\ls}{\lesssim}
\newcommand{\der}[2]{ \frac{ \partial #1 }{\partial #2} }
\newcommand{\cor}[1]{ \left[ #1 \right] }
\newcommand{\pr}[1]{ \left( #1 \right) }

\shorttitle{The influence of thermal evolution in the magnetic protection of terrestrial planets}

\shortauthors{Zuluaga, Bustamante, Cuartas, Hoyos}

\begin{document}

%%%%%%%%%%%%%%%%%%%%%%%%%%%%%%%%%%%%%%%%%%%%%%%%%%%%%%%%%%%%%%%%%%%%%%%%%%%%%%%%%
%FRONT MATTER
%%%%%%%%%%%%%%%%%%%%%%%%%%%%%%%%%%%%%%%%%%%%%%%%%%%%%%%%%%%%%%%%%%%%%%%%%%%%%%%%%

\title{The influence of thermal evolution in the magnetic \\ protection
  of terrestrial planets}

\author{Jorge I. Zuluaga\altaffilmark{1}, Sebastian
  Bustamante\altaffilmark{2}, Pablo A. Cuartas\altaffilmark{3}}

\affil{Instituto de F\'isica - FCEN, Universidad de Antioquia,\\ Calle
  67 No. 53-108, Medell\'in, Colombia} 

\and

\author{Jaime H. Hoyos\altaffilmark{4}}
\affil{Departamento de Ciencias B\'asicas, Universidad de
        Medell\'in,\\ Carrera 87 No. 30-65, Medell\'in, Colombia}

\altaffiltext{1}{jzuluaga@fisica.udea.edu.co}
\altaffiltext{2}{sbustama@pegasus.udea.edu.co}
\altaffiltext{3}{p.cuartas@fisica.udea.edu.co}
\altaffiltext{4}{jhhoyos@udem.edu.co}

%%%%%%%%%%%%%%%%%%%%%%%%%%%%%%%%%%%%%%%%%%%%%%%%%%%%%%%%%%%%%%%%%%%%%%%%%%%%%%%%%
%ABSTRACT
%%%%%%%%%%%%%%%%%%%%%%%%%%%%%%%%%%%%%%%%%%%%%%%%%%%%%%%%%%%%%%%%%%%%%%%%%%%%%%%%%

\begin{abstract}
Magnetic protection of potentially habitable planets plays a central
role in determining their actual habitability and/or the chances of
detecting atmospheric biosignatures.  We develop here a thermal
evolution model of potentially habitable Earth-like planets and
super-Earths.  Using up-to-date dynamo scaling laws we predict the
properties of core dynamo magnetic fields and study the influence of
thermal evolution on their properties.  The level of magnetic
protection of tidally locked and unlocked planets is estimated by
combining simplified models of the planetary magnetosphere and a
phenomenological description of the stellar wind.  Thermal evolution
introduces a strong dependence of magnetic protection on planetary
mass and rotation rate.  Tidally locked terrestrial planets with an
Earth-like composition would have early dayside magnetospause
distances between 1.5 and 4.0 $R_p$, larger than previously estimated.
Unlocked planets with periods of rotation $\sim 1$ day are protected
by magnetospheres extending between 3 and 8 $R_p$.  Our results are
robust against variations in planetary bulk composition and
uncertainties in other critical model parameters.  For illustration
purposes the thermal evolution and magnetic protection of the
potentially habitable super-Earths GL 581d, GJ 667Cc and HD 40307g were
also studied. Assuming an Earth-like composition we found that the
dynamos of these planets are already extinct or close to being shut
down.  While GL 581d is the best protected, the protection of HD
40307g cannot be reliably estimated.  GJ 667Cc, even under optimistic
conditions, seems to be severely exposed to the stellar wind and,
under the conditions of our model, has probably suffered massive
atmospheric losses.
% \vspace{0.3cm}
\end{abstract}
%REVISED2-MAN

%%%%%%%%%%%%%%%%%%%%%%%%%%%%%%%%%%%%%%%%%%%%%%%%%%%%%%%%%%%%%%%%%%%%%%%%%%%%%%%%%
%KEY WORDS
%%%%%%%%%%%%%%%%%%%%%%%%%%%%%%%%%%%%%%%%%%%%%%%%%%%%%%%%%%%%%%%%%%%%%%%%%%%%%%%%%


%%%%%%%%%%%%%%%%%%%%%%%%%%%%%%%%%%%%%%%%%%%%%%%%%%%%%%%%%%%%%%%%%%%%%%%%%%%%%%%%%
%PAPER CONTENT
%%%%%%%%%%%%%%%%%%%%%%%%%%%%%%%%%%%%%%%%%%%%%%%%%%%%%%%%%%%%%%%%%%%%%%%%%%%%%%%%%

%%%%%%%%%%%%%%%%%%%%%%%%%%%%%%%%%%%%%%%%%%%%%%%%%%%%%%%%%%%%%%%%%%%%%%%%%
\section{Introduction}
\label{sec:introduction}
%%%%%%%%%%%%%%%%%%%%%%%%%%%%%%%%%%%%%%%%%%%%%%%%%%%%%%%%%%%%%%%%%%%%%%%%%

The discovery of extrasolar habitable planets is one of the most
ambitious challenges in exoplanetary research.  At the time of
writing, there are almost 861 confirmed exoplanets\footnote{For
  updates, please refer to {http://exoplanet.eu}} including 61
classified as Earth-like planets (EPs, $M\sim 1\,M_\oplus$) and
super-Earths (SEs, $M\sim\,1-10\,M_\oplus$, \citealt{Valencia06} hereafter VAL06).
Among these low mass planets there are three confirmed SEs, GJ 667Cc
\citep{Bonfils11b}, GL 581d \citep{Udry07,Mayor09}, and HD 40307g
\citep{Tuomi2012}, and tens of Kepler candidates \citep{Borucki11,
  Batalha12} that are close or inside the habitable zone (HZ) of their
host stars (see e.g. \citealt{Selsis07a,Pepe11,Kaltenegger11a}).  If
we include the possibility that giant exoplanets could harbour
habitable exomoons, the number of the already discovered potentially
habitable planetary environments beyond the Solar System could be
rised to several tens \citep{Underwood03,Kaltenegger10a}.  Moreover,
the existence of a plethora of other terrestrial planets (TPs) and
exomoons in the Galaxy is rapidly gaining evidence
\citep{Borucki11,Catanzarite11,Bonfils11b,Kipping12}, and the chances
that a large number of potentially habitable extrasolar bodies could
be discovered in the near future are very large.
%REVISED-MAN

The question of which properties a planetary environment needs in
order to allow the appearance, evolution and diversification of life
has been extensively studied (for recent reviews see
\citealt{Lammer09} and \citealt{Kasting10}).  Two basic and
complementary physical conditions must be fulfilled: the presence of
an atmosphere and the existence of liquid water on the surface 
\citep{Kasting93}.  However, the fulfilment of these basic conditions
depends on many complex and diverse endogenous and exogenous factors
(for a comprehensive enumeration of these factors see
e.g. \citealt{Ward00} or \citealt{Lammer10})
%REVISED-MAN

The existence and long-term stability of an intense planetary magnetic
field (PMF) is one of these relevant factors (see
e.g. \citealt{Griebmeier10} and references therein).  It has been
shown that a strong enough PMF could protect the atmosphere of
potentially habitable planets, especially its valuable content of
water and other volatiles, against the erosive action of the stellar
wind \citep{Lammer03,Lammer07,Khodachenko07,Chaufray07}.  Planetary
magnetospheres also act as shields against the potentially harmful
effects that the stellar high energy particles and galactic cosmic
rays (CR) produce in the life-forms evolving on the planetary surface
(see e.g. \citealt{Griebmeier05}).  Even in the case that life could
arise and evolve on unmagnetized planets, the detection of atmospheric
biosignatures would be also affected by a higher flux of high energy
particles including CR, especially if the planet is close to very
active M-dwarfs (dM) \citep{Grenfell07,Segura10}.
%REVISED-MAN

It has been recently predicted that most of the TPs in our Galaxy
could be found around dM stars \citep{Boss06, Mayor08, Scalo07,
  Rauer11, Bonfils11b}. Actually $\sim 20\%$ of the presently
confirmed super-Earths belong to planetary systems around stars of
this type.  Planets in the HZ of low mass stars ($\Ms \lesssim 0.6
\MSUN$) would be tidally locked \citep{Joshi97,Heller11}, a condition
that poses serious limitations to their potential habitability (see
e.g. \citealt{Kite11} and references therein).  Tidally locked planets
inside the HZ of dMs have periods in the range of $5-100$ days, a
condition that has commonly been associated with the almost complete
lack of a protective magnetic field \citep{Griebmeier04}.  However,
the relation between rotation and PMF properties, that is critical at
assessing the magnetic protection of slowly rotating planets, is more
complex than previously thought.  In particular, a detailed knowledge
of the thermal evolution of the planet is required to predict not only
the intensity but also the regime (dipolar or multipolar) of the PMF
for a given planetary mass and rotation rate \citep{Zuluaga12}.
%REVISED-MAN

Although several authors have extensively studied the protection that
intrinsic PMF would provide to extrasolar planets \citep{Griebmeier05,
  Khodachenko07, Lammer07, Griebmeier09, Griebmeier10}, all have
disregarded the influence that thermal evolution has in the evolution
of planetary magnetic properties.  They have used also outdated dynamo
scaling-laws that have been recently revised (see
\citealt{Christensen10} and references therein).  The role of rotation
in determining the PMF properties that is critical in assessing the
case of tidally locked planets has also been overlooked
\citep{Zuluaga12}.
%REVISED-MAN

We develop here a comprenhensive model for the evolution of the
magnetic protection of potentially habitable TPs around GKM main
sequence stars.  To achieve this goal we integrate in a single
framework a parametrized thermal evolution model based on the most
recent advances in the field
\citep{Gaidos10,Tachinami11,Stamenkovic12}, up-to-date dynamo
scaling-laws \citep{Christensen10,Zuluaga12} and phenomenological
models for the evolution of the stellar wind and planet-star magnetic
interaction \citep{Griebmeier10}.  Our model is aimed at: 1)
understanding the influence of thermal evolution in the magnetic
protection of TPs, 2) assesing the role of low rotation periods in the
evolution of the magnetic protection of tidally-locked habitable
planets, 3) placing more realistic constraints on the magnetic
properties of potentially habitable TPs suitable for future studies of
atmospheric mass-loss or the CR effect on the atmospheric chemistry or
on life itself and 4) estimating by the first time the magnetic
properties of already discovered super-Earths in the HZ of their host
stars.
%REVISED-MAN

Our work is a step-forward in the understanding of planetary magnetic
protection because it puts together in a single model the evolution of
the magnetic properties of the planet and hence its dependence on
planetary mass and composition and the role of the planet-star
interaction into determining the resulting level of magnetic
protection.  Previous models of the former (thermal evolution and
intrinsic magnetic properties) did not consider the interaction of the
planetary magnetic field with the evolving stellar wind which is
finally the factor that determine the actual level of planetary
magnetic protection.  On the other hand previous attempts to study the
planet-star interaction overlooked the non-trivial dependence of
intrinsic magnetic properties on planetary thermal evolution and hence
on planetary mass and rotation rate.
%% two relevant aspects of the problem: 1) the dependence of magnetic
%% properties on the thermal evolution of the planet and hence on
%% planetary mass and composition and 2) the role of rotation into the
%% determination of magnetic field intensity and regime.
Additionally, and for the first time, we are attempting here to
calculate the magnetic properties of the already discovered
potentially habitable super-Earths, GJ 667Cc, Gl 581d and HD 40307g.
%REVISED-MAN

This paper is organized as follows.  Section
\ref{sec:MagnetosphereModel} is aimed at introducing the properties of
planetary magnetospheres we should estimate in order to evaluate the
level of magnetic protection of a potentially habitable terrestrial
planet.  Once those properties are expressed in terms of two basic
physical quantities, the planetary magnetic dipole moment and the
pressure of the stellar wind, we proceed to describe how those
quantities can be estimated by modelling the thermal evolution of the
planet (section \ref{sec:ThermalEvolution}), scaling the dynamo
properties from the planetary thermal and rotational properties
(section \ref{sec:ScalingPMF}), and modelling the interaction between
the star and the planet (section \ref{sec:PlanetStarInteraction}).  In
section \ref{sec:Results} we apply our model to evaluate the level of
magnetic protection of hypothetical potentially habitable TPs as well
as the already discovered habitable super-Earths.  We also present
there the results of a numerical analysis aimed at evaluating the
sensitivity of our model to uncertainties in the composition of the
planet and to other critical parameters of the model.  In section
\ref{sec:Discussion} we discuss the limitations of our model, present
an example of the way in which our results could be applied to
estimate the mass-loss rate from already and yet to be discovered
potentially habitable TPs and discuss the observational prospects to
validate or improve the model.  Finally a summary and several
conclusions drawn from this research are presented in section
\ref{sec:Conclusions}.
%REVISED-MAN

%%%%%%%%%%%%%%%%%%%%%%%%%%%%%%%%%%%%%%%%%%%%%%%%%%%%%%%%%%%%%%%%%%%%%%%%%
\section{Critical properties of an evolving magnetosphere}
\label{sec:MagnetosphereModel}
%%%%%%%%%%%%%%%%%%%%%%%%%%%%%%%%%%%%%%%%%%%%%%%%%%%%%%%%%%%%%%%%%%%%%%%%%

The interaction between the PMF, the interplanetary magnetic field
(IMF) and the stellar wind creates a magnetic cavity around the planet
known as the magnetosphere.  Although magnetospheres are very complex
systems, their global properties are continuous functions of only two
physical variables \citep{Siscoe75b}: the magnetic dipole moment of
the planet, ${\cal M}$, and the dynamic pressure of the stellar
wind, $P_\rom{sw}$.  Dipole moment is defined in the multipolar
expansion of the magnetic field strength:
%REVISED-MAN

\beq{eq:M}
B_{p}(r)=\frac{\mu_0{\cal M}}{4\pi r^3}+{\cal
  O}\left(\frac{1}{r^4}\right),
\eeq
%REVISED-MAN

where $B_{p}(r)$ is the angular-averaged PMF strength measured at a
distance $r$ from the planet center and $\mu_0=4 \pi \times
10^{-7}\,\rm{H/m}$ is the vacuum permeability.  In the rest of the
paper we will drop-off the higher order terms in $1/r$ (multipolar
terms) and focus on the dipolar component of the field $B_p^{\rm dip}$
which is explicitly given by the first term of the right side in
eq. (\ref{eq:M}).
%REVISED-MAN

The dynamic pressure of the stellar wind is given by

\beq{eq:Psw}
P_\rom{sw}=m n v_\rom{eff}^2+2 n k_B T.
\eeq
%REVISED-MAN

where $m$ and $n$ are the typical mass of a wind particle (mostly
protons) and its number density, respectively.  Here
$v_\rom{eff}=(v_\rom{sw}^2+v_{p}^2)^{1/2}$ is the effective velocity
of the stellar wind as measured in the reference frame of the planet
whose orbital velocity is $v_p$.  $T$ is the local temperature of the
wind plasma and $k_B=1.38\times 10^{-23}\;{\rm j/K}$ is the Boltzmann
constant.
%REVISED-MAN

There are three basic properties of planetary magnetospheres we are
interested in: 1) the maximum magnetopause field intensity $B_{mp}$,
which is a proxy of the flux of high energy particles entering into
the magnetospheric cavity, 2) the standoff or stagnation radius,
$R_S$, a measure of the size of the dayside magnetosphere, and 3) the
area of the polar cap $A_{pc}$ that measures the total area of the
planetary atmosphere exposed to open field lines through which
particles can escape to interplanetary space.  The value of these
quantities provides information about the level of exposure that a
habitable planet has to the erosive effects of stellar wind and the
potentially harmful effects of the CR.
%REVISED-MAN

The maximum value of the magnetopause field intensity $B_\rom{mp}$ is
estimated from the balance between the magnetic pressure
$P_\rom{mp}=B_\rom{mp}^2/(2 \mu_0)$ and the dynamic stellar wind
pressure $P_\rom{sw}$ (eq. (\ref{eq:Psw})),
%REVISED-MAN

\beq{eq:BmpPsw}
B_\rom{mp}=(2 \mu_0)^{1/2} P_\rom{sw}^{1/2}
\eeq
%REVISED-MAN

Here we are assuming that the pressure exerted by the plasma inside
the magnetospheric cavity is negligible (see discussion below).
%REVISED-MAN

Although magnetopause fields arise from very complex processes
(Chapman-Ferraro and other complex currents at the magnetosphere
boundary), in simplified models $B_\rom{mp}$ is assumed proportional
to the PMF intensity $B_\rom{p}$ as measured at the substellar point
$r=R_S$ \citep{Mead64,Voigt95},
%REVISED-MAN

\beq{eq:BmpMRs}
B_\rom{mp}=2 f_0 B_\rom{p}(r=R_S)\approx \left(\frac{f_0
  \mu_0}{2\pi}\right) \sqrt{2} {\cal M} R_S^{-3}
\eeq
%REVISED-MAN

$f_0$ is a numerical enhancement factor of order 1 that can be
estimated numerically.  We are assuming here that the dipolar
component of the intrinsic field (first term in the r.h.s. of
eq. (\ref{eq:M})) dominates at magnetopause distances even in slightly
dipolar PMF.
%REVISED-MAN

Combining equations (\ref{eq:BmpPsw}) and (\ref{eq:BmpMRs}) we
estimate the standoff distance:
%REVISED-MAN

$$
R_S  =  \left(\frac{\mu_0 f_0^2 }{8 \pi^2}\right)^{1/6}
\mathcal{M}^{1/3} P_\rom{sw}^{-1/6}
$$

that can be expressed in terms of the present dipole moment of the
Earth ${\cal M}_\oplus = 7.768\times 10^{22}$ A m$^2$ and the average
dynamic pressure of the solar wind as measured at the orbit of our
planet $P_\rom{sw\odot}=2.24\times 10^{-9}$
\citep{Stacey92,Griebmeier05}:

\beq{eq:Rs}
\frac{R_S}{R_\oplus} = 9.75 \left(\frac{\cal
  M}{{\cal M}_\oplus}\right)^{1/3}
\left(\frac{P_\rom{sw}}{P_\rom{sw\odot}}\right)^{-1/6}
\eeq
%REVISED-MAN
 
It is important to stress that the value of $R_S$ estimated with
eq. (\ref{eq:Rs}) assumes a negligible value of the plasma pressure
inside the magentospheric cavity.  This approximation is valid if at
least one of these conditions is fulfilled: 1) the planetary magnetic
field is very intense, 2) the dynamic pressure of the stellar wind
is small, or 3) the planetary atmosphere is not too bloated by the XUV
radiation.  In the case when any of these conditions are fulfilled we
will refer to $R_S$ as given by eq. (\ref{eq:Rs}) as the {\it magnetic
  standoff distance} which is an underestimation of the actual size of
the magnetosphere.
%REVISED-MAN

Last but not least we are interested in evaluating the area of the
polar cap.  This is the region in the magnetosphere where magnetic
field lines could be open into the interplanetary space or to the
magnetotail region.  \citet{Siscoe75a} have shown that the area of the
polar cap $A_\rom{pc}$ scales with dipole moment and dynamic
pressure as:
%REVISED-MAN

\beq{eq:Apc}
\frac{A_\rom{pc}}{4\pi R_p^2}=4.63\% \left(\frac{\cal
  M}{{\cal M}_\oplus}\right)^{-1/3}
\left(\frac{P_\rom{sw}}{P_\rom{sw\odot}}\right)^{1/6}
\eeq
%REVISED-MAN

Here we have normalized the polar cap area with the total area of the
atmosphere $4\pi R_p^2$, and assumed the atmosphere has a scale-height
much smaller than planetary radius $R_p$.
%REVISED-MAN

In order to model the evolution of these three key magnetosphere
properties we need to estimate the surface dipolar component of the
PMF $B_p^{\rm dip}(R_p)$ (from which we can obtain the dipole moment
${\cal M}$), the average number density $n$, velocity $v_\rom{eff}$
and temperature $T$ of the stellar wind (which are required to predict
the dynamic pressure $P_{\rm sw}$).  These quantities depend in
general on time and also on different planetary and stellar
properties.  In the following sections we describe our model for the
calculation of the evolving values of these fundamental quantities.
%REVISED-MAN

%============================================================
\section{Thermal and dynamo evolution}
\label{sec:ThermalEvolution}
%============================================================

We assume here that the main source of a global PMF in TPs is the
action of a dynamo powered by convection in a liquid metallic core
\citep{Stevenson83, Stevenson03}.  This assumption is reasonable since
the Moon and all rocky planets in the solar system, regardless of
their different origins and compositions, seem to have in the present
or to have had in the past an iron core dynamo (see
e.g. \citealt{Stevenson10}).  Other potential sources of PMFs such as
body currents induced by the stellar magnetic field or dynamo action
in a mantle of ice, water or magma are not considered here but left
for future research.
%REVISED-MAN

The properties and evolution of a core dynamo will depend on the
internal structure and thermal history of the planet.  Thermal
evolution of TPs, specially the Earth itself, has been studied for
decades (for a recent review see \citealt{Nimmo09a}).  A diversity of
thermal evolution models for planets larger than the Earth have
recently appeared in literature \citep{Papuc08, Gaidos10, Tachinami11,
  Driscoll11, Stamenkovic11}.  But the lack of observational evidence
against which we can compare the predictions of these models has left
too much room for uncertainties especially regarding mantle rheology,
core composition and thermodynamic properties.  Albeit these
fundamental limitations, a global picture of the thermal history of
super-Earths has started to arise.  Here we follow the lines of
\citealt{Labrosse03} and \citealt{Gaidos10} developing a parametrized
thermal evolution model which combines a simplified model of the
interior structure and a parametrized description of the core and
mantle rheology.
%REVISED-MAN

Our model includes several distinctive characteristics in comparison
to previous ones.  The most important one is an up-to-date treatment
of mantle rheology.  For that purpose we use two different formulae to
compute the viscosity of the upper and lower mantle.  By lower mantle
we understand here the region of the mantle close to the core mantle
boundary (CMB).  This is the hottest part of the mantle.  The upper
mantle is the outer cold part of this layer.  It is customary to
describe both regions with the same rheology albeit their very
different mineralogical compositions.  Additionally thermal and
density profiles in the mantle following the same prescription as in
the core.
%% With this strategy we avoid using very unphysical assumptions such as
%% that of treaing mantle as an isothermal layer (see
%% e.g. \citealt{Gaidos10}).  
We also use a different ansatz to assign initial values to lower
mantle temperature and to the temperature contrast across the CMB, 
two of the most uncertain quantities in thermal
evolution models.  Using our ansatz we avoid assigning arbitrary
initial values to these critical parameters but more importantly we
are able to find a unified method to set the value of these
temperatures in planets with very different masses.  It is also
important to notice that in other models these temperatures were set
by hand
%(e.g. \citealt{Papuc08}) 
or were treated as free parameters in the model
%(e.g. \citealt{Tachinami11}).
%REVISED-MAN

Four key properties should be predicted by any thermal evolution model
in order to calculate the magnetic properties of a planet: 1) the
total available convective power $Q_\rom{conv}$, providing the energy
required for magnetic field amplification through dynamo action; 2)
the radius of the solid inner core $R_\rom{ic}$ and from there the
height $D\approx R_c-R_\rom{ic}$ of the convecting shell where the
dynamo action takes place ($R_c$ is the radius of the core); 3) the
time of inner core formation $t_\rom{ic}$ and 4) the total dynamo
life-time $t_\rom{dyn}$.
%REVISED-MAN

In order to calculate these quantities we solve simply parametrized
energy and entropy equations of balance describing the flux of heat
and entropy in the planetary core and mantle.  As stated before our
model is based on the interior structure model by VAL06) 
and in thermal evolution models previously developed
by \citealt{Schubert79}, \citealt{Stevenson83}, \citealt{Nimmo00},
\citealt{Labrosse01}, \citealt{Labrosse03}, \citealt{Gubbins03},
\citealt{Gubbins04},
\citealt{Aubert09},\citealt{Gaidos10},\citealt{Stamenkovic11}.  For a
detailed description of the fundamental physics behind the thermal
evolution model developed here please refer to these earlier studies.
%REVISED-MAN

%----------------------------------------------------------------------
\subsection{Interior structure}
\label{subsec:InteriorStructure}
%----------------------------------------------------------------------

Our one-dimensional model for the interior assumes a planet made by
two well differentiated chemically and mineralogically homogeneous
shells: a rocky mantle made out of olivine and perovskite and a core
made by iron plus other light elements.
%REVISED-MAN

The mechanical conditions inside the planet (pressure $P$, density
$\rho$ and gravitational field $g$) are computed by solving
simultaneously the continuity, Adams-Williamson and hydrostatic
equillibrium equations (eqs. (1)-(4) in VAL06).  For all planetary masses
we assume boundary conditions, $\rho(r=R_p)=4000$ kg
m$^{-3}$ and $P(r=R_p)=0$ Pa. For each planetary mass and core mass
fraction CMF$=M_{\rm core}/M_p$ we use a RK4 integrator and a shooting
method to compute consistently the core $R_c$ and planetary radius
$R_p$.
%REVISED-MAN

For the sake of simplicity, we do not include in the interior model
the two- or even three-layered structure of the mantle. Instead of
that we assume a mantle completely made of perovskite-postperovskite
(ppv).  This is the reason why we take $\rho(r=R_p)=4000$ kg m$^{-3}$
instead of the more realistic value of $3000$ kg m$^{-3}$.  With a
single layer and a realistic surface density our model is able to
reproduce the present interior properties of the Earth.

In all cases we use the Vinet equation of state instead of the
commonly used third order Birch-Murnaghan equation (BM3).  It is well
known that the BM3 follows from a finite strain expansion and does not
accurately predict the properties of the material for the typical
pressures found in super-Earths , i.e. $100-1000$ GPa for
$M_p=1-10\ME$ (VAL06 and \citealt{Tachinami11}).  We have ignored
thermal corrections to the adiabatic compressibility,
i.e. $K_S(\rho,T) \approx K_S( \rho, 300\mbox{ K}) + \Delta K_s( T )$
(VAL06).  This assumption allows us to decouple at runtime the CPU
intensive calculation of the thermal profile from the mechanical
structure at each time-step in the thermal evolution integration.
%REVISED-FIN

Although we have ignored the ``first-order'' effect of the temperature
in the mechanical structure, we have taken into account
``second-order'' effects produced by phase transitions inside the
mantle and core.  Using the initial temperature profile inside the
mantle we calculate the radius of transition from olivine to
perovskite (neglecting the effect of an intermediate layer of
wadsleyite).  For that purpose we use the reduced pressure function
$\Pi$ \citep{Christensen85,Weinstein92,Valencia07a}.  For simplicity
the position of the transition layer is assumed constant during the
whole thermal evolution of the planet.  We have verified that this
assumption does not change significantly the mechanical properties
inside the planet at least not at a level affecting the thermal
evolution itself.
%REVISED-FIN

Inside the core we update continuously the radius of the transition
from solid to liquid iron (see below).  For that purpose we use the
thermal profile computed at the previous time-step.  To avoid a
continuous update of the mechanical structure we assume in all the
cases that at the transition from solid to liquid the density of iron
changes by a constant factor $\Delta \rho =
(\rho_{s}-\rho_{l})/\rho_{l}$.  Here the reference density of the
solid $\rho_{s}$ (when applied) is computed using the Vinet equation
evaluated at a point in the very center of the planet.
%REVISED-FIN

Table \ref{tab:ThermalModelParameters} enumerates the relevant
physical parameters of the interior structure and thermal evolution
model.
%REVISED-FIN

%----------------------------------------------------------------------
\subsection{Core thermal evolution}
\label{subsec:CoreThermalEvolution}
%----------------------------------------------------------------------

In order to compute the thermal evolution of the core we solve the
equations of energy and entropy balance \citep{Labrosse01,Nimmo09a}:
%REVISED-MAN

\beq{eq:EnergyBalance}
Q_c = Q_s + f_i( Q_g + Q_l )
\eeq

\beq{eq:EntropyBalance}
\Phi = E_l + f_i( E_s + E_g ) - E_k.
\eeq

Here $Q_c$ is the total heat flowing through the CMB and $\Phi$ is the
total entropy dissipated in the core.  $E_s$ and $Q_s$ are the entropy
and heat released by the secular cooling, $E_g$ and $Q_g$ are the
contribution to entropy and heat due to the redistribution of
gravitational potential when light elements are released at the
liquid-solid interface (buoyant energy), $E_l$ and $Q_l$ are the
entropy and heat released by the phase transition (latent heat) and
$E_k$ is a term accounting for the sink of entropy due to the
conduction of heat along the core.  We have avoided the terms coming
from radioactive and pressure heating because their contribution are
negligible at the typical conditions inside super-Earths
\citep{Nimmo09a}.  As long as the buoyant and latent entropy and heat
are only present when a solid inner core exists we have introduced a
boolean variable $f_i$ that turn-on these terms when the condition for
the solidification of the inner core arises.
%REVISED-MAN

The terms in the energy and entropy balance are a function of the
time-derivative of the temperature profile $\partial T(r,t)/\partial
t$ (for detailed expressions of these terms see table 1 in
\citealt{Nimmo09a}).  As an example the secular heat and entropy are
given by:
%REVISED-MAN

\begin{eqnarray}
\label{eq:SecularHeat}
Q_s & = & -\int\rho c_p\frac{\partial T(r,t)}{\partial t}dV, \nonumber\\
E_s & = & -\int\rho c_p \left[\frac{1}{T_c(t)} - \frac{1}{T(r,t)}\right]\frac{\partial
  T(r,t)}{\partial t}dV.
\end{eqnarray}

Here $c_p$ is the specific heat of the core alloy and $T_c$ is the
temperature at the CMB.  If we assume that the temperature profile of
the core does not change during the thermal evolution, we can write
temperature as $T(r,t) = f_c(r)T_c(t)$.  Here $f_c(r)$ is the core
temperature radial profile that we will assume adiabatic (see below).
It should be noted again that $T_c(t) = T(r=R_c,t)$.

With this assumption the energy balance equation
(\ref{eq:EnergyBalance}) can be written as a first order differential
equation on $T_c$:
%REVISED-MAN

\beq{eq:TEquation}
Q_c=M_c [C_s + f_i( C_g + C_l )] \dtot{T_c}{t}
\eeq

where $M_c$ is the total mass of the core and $C_s$, $C_g$ and $C_l$
are core {\it bulk heat capacities} which can be expressed as
volumetric integrals of the radial profile $f_c(r)$.  In this equation
the total heat $Q_c$ is intrinsically a function of $T_c$ and should
be computed independently (see below).
%REVISED-MAN

Using a simple exponential fit for the core density, as proposed by
\citet{Labrosse01}, the adiabatic temperature profile can be
approximated as \citep{Labrosse03}:
%REVISED-MAN

\beq{eq:CoreTemperatureProfile}
f_c(r) = \exp\left(\frac{R_c^2 - r^2}{D_c^2}\right)
\eeq

where $D_c=\sqrt{3 c_p/2\pi\alpha\rho_c G}$ is the scale height of
temperature, $\alpha$ is the isothermal expansivity (assumed for
simplicity constant along the core) and $\rho_c$ is the density at
core center. Using this fit the bulk secular heat capacity $C_s\equiv
Q_s/(M_c\;dT_c/dt)$ can be obtained from eq. (\ref{eq:SecularHeat}):
%REVISED-MAN

\beq{eq:SecularHeatCapacity}
C_s=-4\pi\int_0^{R_c}\rho(r)c_p\exp\left(\frac{R_c^2 -
  r^2}{D_c^2}\right) r^2 dr
\eeq

Analogous expressions for $C_g$ and $C_l$ are obtained from the
definition of $Q_g$ and $Q_l$ as given in table 1 of \citet{Nimmo09a}.
%REVISED-MAN

The total heat released by the core $Q_c(T_c)$ is calculated here
using the boundary layer theory (BLT) (see
e.g. \citealt{Stevenson83}).  Under this approximation $Q_c$ is given
by \citep{Ricard09}:
%REVISED-MAN

\beq{eq:Qc}
Q_c = 4\pi R_c k_m \Delta T_{\rm CMB} {\rm Nu}_c,
\eeq

where $k_m$ is the thermal conductivity of the lower mantle, $\Delta
T_{\rm CMB}=T_c - T_l$ is the temperature contrast across the CMB,
$T_l$ is the lower mantle temperature, and
Nu$_c\approx(\mbox{Ra}_c/\mbox{Ra}_*)^{1/3}$ is the Nusselt number at
the core \citep{Schubert01}.  The critical Rayleigh number
$\mbox{Ra}_*$ is a free parameter in our model (see table
\ref{tab:ThermalModelParameters}).
%REVISED-MAN

The local Rayleigh number Ra$_c$ at the CMB is calculated under the
assumption of a boundary heated from below \citep{Ricard09},
%REVISED-MAN

\beq{eq:Rac}
\mbox{Ra}_c = \frac{\rho\; g\; \alpha\;\Delta T_{\rm CMB}(R_p -
  R_c)^3}{\kappa_c \eta_c},
\eeq

where $g$ is the gravitational field and $\kappa_c$ the thermal
diffusivity at the CMB.  The value of the dynamic viscosity
$\eta_c$, which is strongly dependent on temperature, could be
suitably computed using the so-called film temperature (see
\citealt{Manga01} and reference therein).  This temperature could be
computed in general as a weighted average of the temperatures at the
boundaries,
%REVISED-FIN

\beq{eq:Tfilm}
T_{{\eta}c}=\xi_cT_c + (1-\xi_c) T_l,
\eeq

where the weighting coefficient $\xi_c$ is a free parameter whose
value is chosen in order to reproduce the thermal properties of the
Earth (see table \ref{tab:ThermalModelParameters}).
%REVISED-FIN

To model the formation and evolution of the solid inner core we need
to compare at each time the temperature profile with the iron solidus.
We use here the Lindemann law as parametrized by VAL06:
%REVISED-FIN

\beq{eq:IronSolidus}
\der{\log \tau}{\log \rho} = 2\cor{ \gamma - \delta(\rho) }.
\eeq

Here $\delta(\rho)\approx1/3$ and $\gamma$ is an effective Gr\"uneisen
parameter that is assumed for simplicity constant.  To integrate this
equation we use the numerical density profile provided by the interior
model and the reference values $\rho_{0} = 8300\mbox{ kg/m}^3$ (pure
iron) and $\tau_0 = 1808$ K.
%REVISED-FIN

The central temperature $T(r=0,t)$ and the solidus at that point
$\tau(r=0)$ are compared at each time step.  When $T(0,t_{\rm
  ic})\approx \tau(0)$ ($t_{\rm ic}$ is the time of inner core
formation) we turn-on the buoyant and latent heat terms in
eqs. (\ref{eq:EnergyBalance}) and (\ref{eq:EntropyBalance}), i.e. set
$f_i=1$, and continue the integration including these terms.  The
radius of the inner core at times $t>t_{\rm ic}$ is obtained by
solving the equation proposed by \citet{Nimmo09a} and further
developed by \citet{Gaidos10},
%REVISED-MAN

\beq{eq:Ric}
\dtot{R_{\rm ic}}{t}=-\frac{D_c^2}{2R_{\rm ic}(\Delta -
  1)}\frac{1}{T_c}\dtot{T_c}{t}.
\eeq

Here $\Delta$ is the ratio between the gradient of the solidus
(eq. (\ref{eq:IronSolidus})) and the actual temperature gradient
$T_c(t)f_c(r)$ as measured at $R_{\rm ic}(t)$.
%REVISED-MAN

When the core cools down below a given level the outer layers start to
stratificate.  Here we model the effect of stratification by
correcting the radius and temperature of the core following the
prescription by \citet{Gaidos10}.  When stratified the effective
radius of the core is reduced to $R_\star$ (eq. (27) in
\citealt{Gaidos10}) and the temperature at the core surface is
increase to $T_\star$ (eq. (28) in \citealt{Gaidos10}).  The
stratification of the core reduces the height of the convective shell
which leads to a reduction of the Coriolis force potentially enhancing
the intensity of the dynamo-generated magnetic field.
%REVISED-MAN

The estimation of the dynamo properties requires the computation of
the available convective power $Q_\rom{conv}$.  $Q_\rom{conv}$ is
calculated here assuming that most of the dissipation occurs at the
top of the core.  Under this assumption, 
%REVISED-MAN

\beq{eq:Qconv}
Q_\rom{conv}(t)\approx\Phi(t) T_c(t)
\eeq

where the total entropy $\Phi$ is computed from the Entropy balance
(eq. (\ref{eq:EntropyBalance})) using the solution for the temperature
profile $T_c(t)f_c(r)$.

When $\Phi(t)$ becomes negative, i.e. $E_l+E_s+E_g<E_k$ in
eq. (\ref{eq:EntropyBalance}), $Q_\rom{conv}$ gets also negative and
convection is not longer efficient to transport energy across the
outer core.  Under this condition the dynamo is shut down.  The
integration stops when this condition is fulfilled at a time we label
as the dynamo life-time $t_{\rm dyn}$.
%REVISED-MAN

%----------------------------------------------------------------------
\subsection{Mantle thermal evolution}
\label{subsec:MantleThermalEvolution}
%----------------------------------------------------------------------

One of the novel features of our thermal evolution model is that we
treat mantle thermal evolution with a similar formalism as that
described before for the metallic core.

The energy balance in the mantle can be written as:
%REVISED-MAN

\beq{eq:MantleEnergyBalance}
Q_m = \chi_r Q_r + Q_s + Q_c
\eeq

Here $Q_m$ is the total heat flowing out through the surface boundary
(SB), $Q_r$ is the heat produced in the decay of radioactive nuclides
inside the mantle, $Q_s$ is the secular heat and $Q_c$ is the heat
coming from the core (eq. (\ref{eq:Qc})).  

We use here the standard expressions and parameters for the
radioactive energy production as given by \citet{Kite09}.  However, in
order to correct for the non-homogeneous distribution of radioactive
elements in the mantle, we introduce a multiplicative correction
factor $\chi_r$.  Here we adopt $\chi_r = 1.253$ that fits well the
Earth properties.  We have verified that the thermal evolution is not
too sensitive to $\chi_r$ and have assumed the same value for all
planetary masses.
%REVISED-FIN

\begin{figure}  
  \centering
   \includegraphics[width=0.8
  \textwidth]{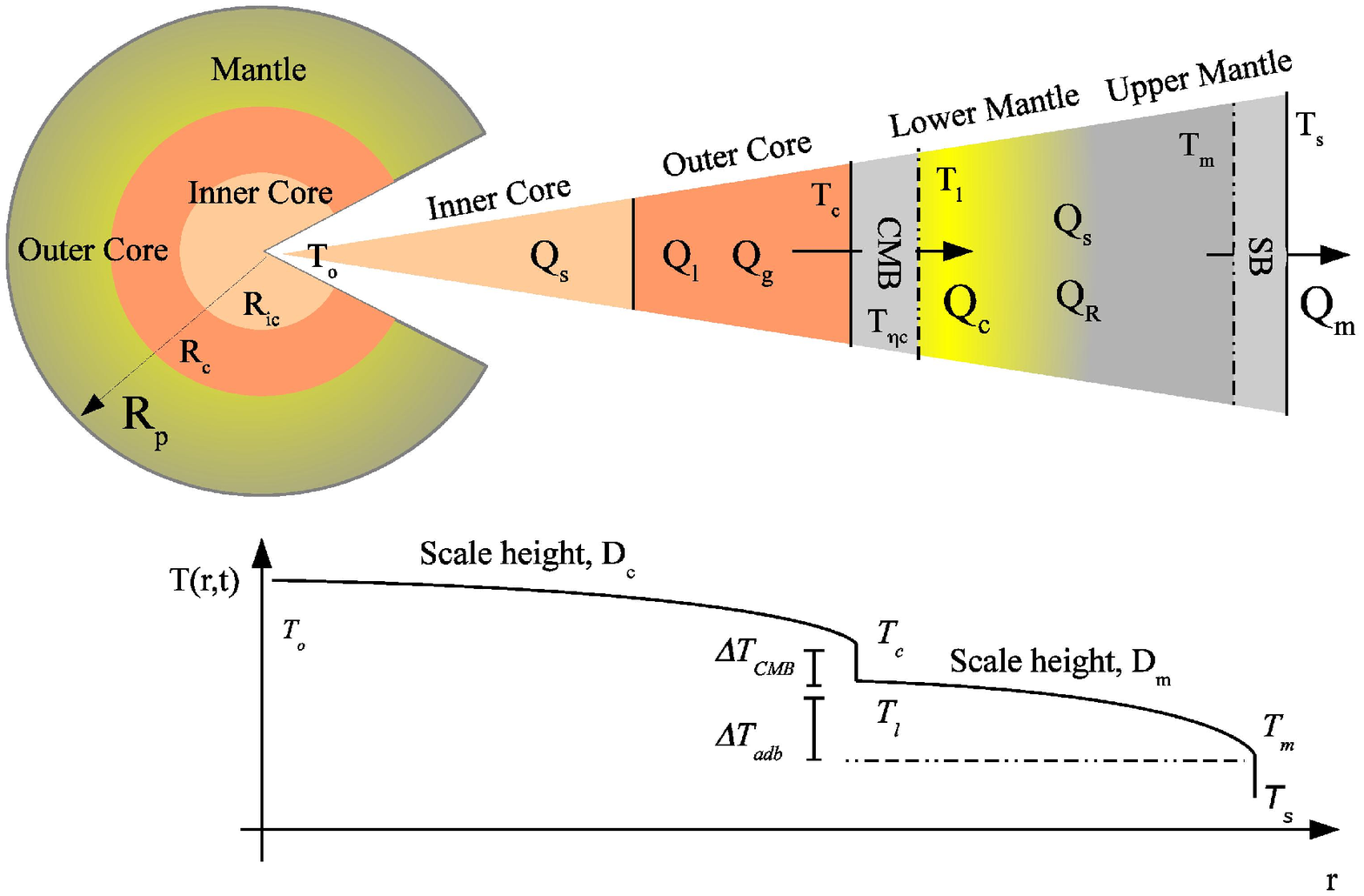}
  \scriptsize
  \caption{Schematic representation of the planetary interior.  In the
    schematic slice we depict the main quantities used here to
    describe the thermal evolution of the planets.  The temperature
    profile depicted below the slice does not use real data.
    Distances and sizes are not represented with the right
    scale.\vspace{0.2cm}} \label{fig:SchematicInterior}
    %REVISED-MAN
\end{figure}

The secular heat in the mantle is computed using an analogous
expression to eq. (\ref{eq:SecularHeat}).  As in the case of the core
we assume that the temperature radial profile does not change during
the thermal evolution.  Under this assumption the temperature profile
in the mantle can be also written as $T(r,t) = T_m(t) f_m(r)$.  In
this case $T_m(t)=T(r=R_p,t)$ is the temperature right below the
surface boundary layer (see figure \ref{fig:SchematicInterior}).
%REVISED-FIN

Assuming an adiabatical temperature profile in the mantle we can
also write:

%\beq{eq:MantleTemperatureProfile}
$$ 
f_m(r) = \exp\left(\frac{R_p^2 - r^2}{D_m^2}\right),
$$

in analogy to the core temperature profile, 
eq. (\ref{eq:CoreTemperatureProfile}).  In this expression $D_m$ is
the temperature scale height for the mantle which is related to the
density scale height $L_m$ through $D_m^2=L_m^2/\gamma$
\citet{Labrosse03}.  In our simplified model we take the values of the
density at the boundaries of the mantle and obtain analytically an
estimate for $L_m$ and hence for $D_m$. 

The energy balance in the mantle is balanced when we independently
calculate the heat $Q_m$ at the SB as a function of $T_m$.  In this
case the presence or not of mobile lids play an important role into
determining the efficiency with which the planet gets rid of the heat
coming from the mantle.  In the mobile lid regime we assume that the
outer layer is fully convective and use the BLT approximation to
calculate $Q_m$,
%REVISED-MAN

\beq{eq:MantleHeat}
Q_{m}^{\rm ML} = \frac{4 \pi R_p^2 k_m \Delta T_m \mbox{Nu}_m}{R_p - R_c}
\eeq

where $\Delta T_m = T_m - T_s$ is the temperature contrast across the
SB and $T_s$ the surface temperature.  Since we are studying the
thermal evolution of habitable planets we assume in all cases
$T_s=290$ K.  Planetary interior structure and thermal evolution are
not too sensitive to surface temperature.  We have verified that
results are nearly the same for surface temperature in the range of
$250-370$ K. In the mobile lid regime $\mbox{Nu}_m$ obeys the same
relationship with the critical Rayleigh number as in the core.  In
this case however we compute the local Rayleigh number under the
assumption of material heated from inside \citep{Gaidos10},
%REVISED-MAN

\beq{eq:Ram}
\mbox{Ra}_m = \frac{\alpha g \rho^2 H (R_p - R_c)^5}{k_m \kappa_m \eta_m}
\eeq

where $H = (Q_r + Q_c)/M_m$ is the density of heat inside the mantle,
$k_m$ and $\kappa_m$ are the thermal conductivity and diffusivity
respectively and $\eta_m$ is the upper mantle viscosity.
%REVISED-MAN

In the stagnant lid regime the SB provides a rigid boundary for the
heat flux.  In this case we adopt the approximation used by
\citealt{Nimmo00}:
%REVISED-MAN

\beq{eq:MantleHeat_Stagnant}
Q_{m}^{\rm SL} = 4\pi R_p^2 \frac{k_m}{2}\pr{ \frac{\rho g \alpha}{\kappa_m 
    \eta_m}}^{1/3}\Gamma^{-4/3}
\eeq

Here $\Gamma \equiv -\partial\ln \eta_m / \partial T_m$ measures the
viscosity dependence on temperature evaluated at the average mantle
pressure.
%REVISED-MAN

With all this elements at hand the energy balance at
eq. (\ref{eq:MantleEnergyBalance}) is finally transformed into an
ordinary differential equation for the upper mantle temperature
$T_m(t)$,
%REVISED-MAN

\beq{eq:TMEquation}
Q_m = \chi_r Q_r + Q_c + C_m \dtot{T_m}{t} 
\eeq

where $C_m$ is the bulk heat capacity of the upper mantle which is
calculated with an analogous expression to
eq. (\ref{eq:SecularHeatCapacity})
%REVISED-MAN

%----------------------------------------------------------------------
\subsection{Initial conditions}
\label{subsec:InitialConditions}
%----------------------------------------------------------------------

In order to solve the coupled differential eqs. (\ref{eq:TEquation}),
(\ref{eq:Ric}) and (\ref{eq:TMEquation}) we need to choose a proper
set of initial conditions.
%REVISED-MAN

The initial value of the upper mantle temperature is chosen using the
prescription by \citet{Stamenkovic11}.  According to this prescription
$T_m(t=0)$ is computed by integrating the pressure-dependent adiabatic
equation up to the average pressure inside the mantle $\langle
P_m\rangle$,
%REVISED-MAN

\beq{eq:PotentialTeperature}
T_{m}(t=0) = \theta
\exp\left(\int_0^{\langle P_m\rangle}\frac{\gamma_0}{K_s(P')}dP'\right)
\eeq

Here $\theta=1700$ K is a potential temperature which is assumed the
same for all planetary masses \citep{Stamenkovic11}.  Using $T_m(t=0)$
and the adiabatic temperature profile we can get the initial lower
mantle temperature $T_l(t=0)$.
%REVISED-MAN

The initial value of core temperature $T_c(t=0)$ is one of the most
uncertain parameters in thermal evolution models.  Although nobody
knows its actual value or its dependence on the formation history and
planetary mass, it is reasonable to start the integration of a
simplified thermal evolution model when the core temperature is of the
same order as the melting point for MgSiO$_3$ at the lower mantle
pressure.  A small arbitrary temperature contrast against this
reference value \citep{Gaidos10,Tachinami11} or more complicated
mass-dependent assumptions \citep{Papuc08} have been used in previous
models to set the initial core temperature.  We use here a simple
prescription that agrees reasonably well with previous attempts and
provides a unified expression that could be used consistently for all
planetary masses.
%REVISED-MAN

According to our prescription the temperature contrast across the CMB
is assumed proportional to the temperature contrast across the whole
mantle, i.e. $\Delta T_{\rm CMB}=\epsilon_{\rm adb} \Delta T_{\rm
  adb}=\epsilon_{\rm adb}(T_m-T_l)$.  We have found that the thermal
evolution properties of Earth are reproduced when we set
$\epsilon_{\rm adb}=0.7$.

Using this prescription the initial core temperature is finally
calculated using:
%REVISED-MAN

\beq{eq:InitialCoreTemperature}
T_c(t=0)=T_l+\epsilon_{\rm adb}\Delta T_{\rm adb}
\eeq

We have observed that the value of the $T_c(t=0)$ obtained with this
prescription is very close to the perovskite melting temperature at
the CMB for all the planetary masses studied here.  This result shows
that although our criterium is not particularly better physically
rooted than those used in previous models, it still relies in just one
free parameter, i.e. the ratio of mantle and CMB contrasts
$\epsilon_{\rm adb}$.

%% Although there is not a simple physical justification for this
%% assumption, it could be partially understood by thinking that the heat
%% released by the core at a temperature gradient $\Delta T_{\rm CMB}$
%% should be also transported through the mantle at a similar temperature
%% contrast $\Delta T_{\rm adb}$.
%REVISED-MAN

%----------------------------------------------------------------------
\subsection{Rheological model}
\label{subsec:Viscosity}
%----------------------------------------------------------------------

One of the most controversial aspects and probably the largest source
of uncertainties in thermal evolution models is the calculation of the
rheological properties of sillicates and iron at high pressures and
temperatures.  A detailed discussion on this important topic is out of
the scope of this paper.  An up-to-date discussion and analysis of the
dependence on pressure and temperature of viscosity in super-Earths
and its influence in thermal evolution can be found in the recent
works by \citealt{Tachinami11} and \citealt{Stamenkovic11}.
%REVISED-MAN

We use here two different models to calculate viscosity under
different ranges of temperatures and pressures.  For the high
pressures and temperatures of the lower mantle we use a
Nabarro-Herring model \citep{Yamazaki01},
%REVISED-MAN

\beq{eq:ViscosityModelLower}
\eta_{\rm NH}(P,T) = \frac{R_gd^m}{D_0 A
  m_{mol}}T\rho(P,T)\exp\pr{\frac{b\;T_{\rm melt}(P)}{T}}
\eeq

Here $R_g = 8.31 \mbox{ Jmol}^{-1}\mbox{K}^{-1}$ is the gas constant,
$d$ is the grain size and $m$ the growing exponent, $A$ and $b$ are
free parameters, $D_0$ is the pre-exponential diffusion coefficient
and $m_{mol}$ is the molar density of perovskite and $T_{\rm melt}(P)$
is the melting temperature of perovskite that can be computed with the
empirical fit:

%\beq{eq:ViscosityModelLower}
$$
T_{\rm melt}(P) = \sum_{i=0}^4 a_i \cdot P^i 
$$
%\eeq

All the parameters used in the visocisty model, including the
expansion coefficients $a_i$ in the melting temperature formula, were
taken from the recent work by \citet{Stamenkovic11}.  The
Nabarro-Herring formula allows us to compute $\eta_c=\eta_{\rm
  NH}(T_{\eta c})$, where $T_{\eta c}$ is the film temperature
computed using the average in eq. (\ref{eq:Tfilm}).
%REVISED-MAN

The upper mantle has a completely different mineralogy and it is under
the influence of lower pressures and temperatures.  Although previous
works have used the same model and parameters to calculate viscosity
across the whole mantle (\citealt{Stamenkovic11} for example use the
perovskite viscosity parameters also in the olivine upper mantle), we
have found here that using a different rheological model in the upper
and lower mantle avoids an under and overestimation, respectively, of
the value of viscosity that could have a significant effect on the
thermal evolution.
%REVISED-MAN

In the upper mantle we find that using an Arrhenius-type model leads
to better estimates of viscosity than that obtained using the
Nabarro-Herring model.  In the upper mantle the Nabarro-Herring
formula (which is best suited to describe the dependence on viscosity
at high-pressures and temperatures) leads to huge underestimations of
viscosity in that region.  In the case of the Earth this
underestimation produces values of the total mantle too high as
compared to that observed in our planet making impossible to fit the
thermal evolution of a simulated Earth.

For the Arrhenius-type formula we use the same parametrization given
by \citet{Tachinami11}:
%REVISED-MAN

\beq{eq:ViscosityModelUpper}
\eta_A(P,T) = \frac{1}{2}\cor{ \frac{1}{B^{1/n}}\exp\pr{ \frac{E^* +
      PV^*}{nR_gT} } }\dot{\epsilon}^{(1-n)/n}
\eeq

where $\dot{\epsilon}$ is the strain rate, $n$ is the creep index, $B$
is the Barger coefficient, and $E^*$ and $V^*$ are the activation
energy and volume.  The values assumed here for these parameters are
the same as that given in table 4 of \citealt{Tachinami11} except for
the activation volume whose value we assume here $V^*=2.5\times
10^{-6}$ m$^3$ mol$^{-1}$.  Using the formula in
eq. (\ref{eq:ViscosityModelUpper}), the upper mantle viscosity is
computed as $\eta_m = \eta_{A}(\langle P_m \rangle, T_m)$.
%REVISED-MAN

\medskip

A summary of the parameters used by our interior and thermal evolution
models are presented in table \ref{tab:ThermalModelParameters}.  The
values listed in column 3 define what we will call the {\it reference
  thermal evolution model} (RTEM).  These reference values have been
mostly obtained by fitting the present interior properties of the
Earth and the global features of its thermal,
dynamo and magnetic field evolution (time of inner core formation and
present values of $R_{\rm ic}$ , $Q_m$ and surface magnetic field
intensity). For the stagnant lid case we use as suggested by
\citealt{Gaidos10} the values of the parameters that globally
reproduce the present thermal and magnetic properties of Venus.
%REVISED-MAN

\begin{figure}  
  \centering
   \includegraphics[width=0.7
  \textwidth]{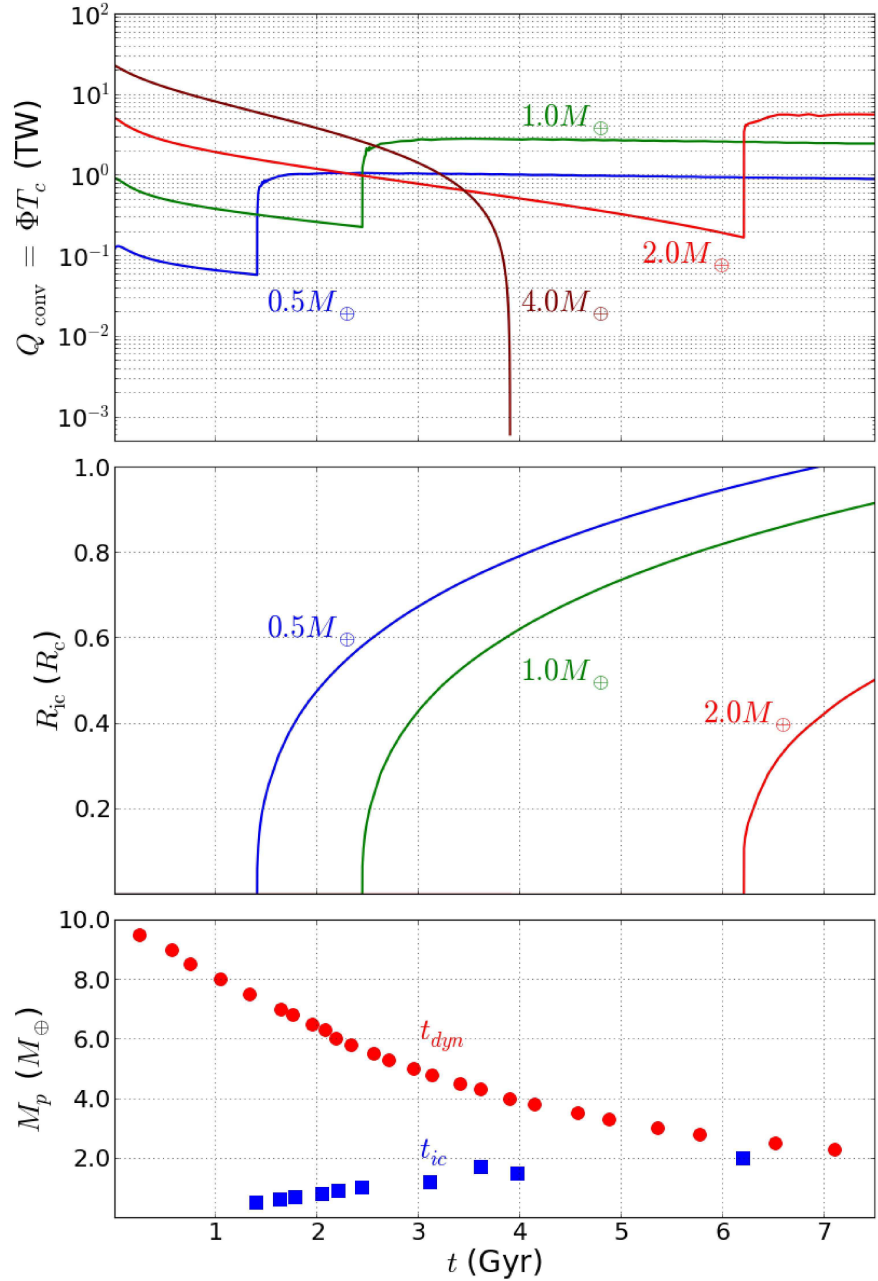}
  \scriptsize
  \caption{Thermal evolution of TPs with an Earth-like composition
    ($\rom{CMF}=0.325$) using the RTEM (see table
    \ref{tab:ThermalModelParameters}).  Upper panel: convective
    power flux $Q_{\rm conv}$ (see eq. (\ref{eq:Qconv})).  
    Middle panel: radius of the
    inner core $R_{ic}$.  Lower panel: time of inner core formation
    (blue squares) and dynamo lifetime (red circles).  In the
    RTEM the metallic core is liquid at $t=0$ for all planetary
    masses.  Planets with a mass $M_p<M_\rom{crit}=2.0\ME$ develop a
    solid inner core before the shut down of the dynamo while the core
    of more massive planets remains liquid at least until the dynamo
    shutdown.\vspace{0.2cm}} \label{fig:ThermalEvolution}
    %REVISED-MAN
\end{figure}

Figure \ref{fig:ThermalEvolution} shows the results of applying the
RTEM to a set of hypothetical TPs in the mass-range $M_p=0.5-6\ME$.
%REVISED-MAN

%TTTTTTTTTTTTTTTTTTTTTTTTTTTTTTTTTTTTTTTTTTTTTTTTTTTTTTTTTTTTTTTTTTTTTTTTTTT
%TABLE 1: THERMAL EVOLUTION PARAMETERS
\begin{table*}
  \centering
  \scriptsize
  \begin{tabular}{llllcc}
    Parameter & Definition & Value & Ref.\\\hline

    \hline\multicolumn{4}{c}{Bulk}\\\hline 
    CMF & Core mass fraction & 0.325 & -- \\
    $T_s$ & Surface temperature & 290 K & -- \\
    $P_s$ & Surface pressure & 0 bar & -- \\

    \hline\multicolumn{4}{c}{Inner core}\\\hline 
    -- & Material & Fe & A \\
    $\rho_0$, $K_0$, $K_0'$, $\gamma_0$, $q$, $\theta_0$ & Equation of state parameters & 8300 kg m$^{-3}$, 160.2 GPa, 5.82, 1.36, 0.91, 998 K & A \\ 
    $k_c$ & Thermal conductivity & 40 W m$^{-1}$ K$^{-1}$ & B \\
    $\Delta S$ & Entropy of fusion & 118 j kg$^{-1}$K$^{-1}$& C \\
    
    \hline\multicolumn{4}{c}{Outer core}\\\hline
    --  & Material  & Fe$_{(0,8)}$FeS$_{(0,2)}$ & A \\
    $\rho_0$, $K_0$, $K_0'$, $\gamma_0$, $q$, $\theta_0$ & Equation of state parameters & 7171 kg m$^{-3}$, 150.2 GPa, 5.675, 1.36, 0.91, 998 K & A \\ 
    $\alpha$ & Thermal expansivity & 1.4 $\times 10^{-6}$ K$^{-1}$ & D \\
    $c_p$ & Specific heat & 850 j kg$^{-1}$ K$^{-1}$& C \\
    $k_c$ & Thermal conductivity & 40 W m$^{-1}$ K$^{-1}$ & B \\
    $\kappa_c$ & Thermal diffusivity & $6.5\times 10^{-6}$ m$^2$ s$^{-1}$ & E \\
    $\Delta S$ & Entropy of fusion & 118 j kg$^{-1}$ K$^{-1}$& C \\
    $\epsilon_{adb}$ & Adiabatic factor for $T_c(t=0)$ & 0.7 & -- \\
    $\xi_c$  & Weight of $T_c$ in core viscosity & 0.4 & -- \\

    \hline\multicolumn{4}{c}{Lower mantle}\\\hline
    -- & Material  & pv+fmw & A \\
    $\rho_0$, $K_0$, $K_0'$, $\gamma_0$, $q$, $\theta_0$ & Equation of state parameters & 4152 kg m$^{-3}$, 223.6 GPa, 4.274, 1.48, 1.4, 1070 K & A \\ 
    $d$, $m$, $A$, $b$, $D_0$, $m_{mol}$ & Viscosity parameters & 1$\times 10 ^{-3}$ m, 2, 13.3, 12.33, 2.7$\times 10^{-10}$ m$^2$ s$^{-1}$, 0.10039 kg mol$^{-1}$ & F \\
    $\alpha$ & Thermal expansivity & 2.4 $\times 10^{-6}$K$^{-1}$ & D \\
    $c_p$ & Specific heat & 1250 j kg$^{-1}$ K$^{-1}$& C \\
    $k_m$ & Thermal conductivity & 6 W m$^{-1}$ K$^{-1}$ & C \\
    $\kappa_m$ & Thermal diffusivity & $7.5\times 10^{-7}$ m$^2$s$^{-1}$ & E \\
    $\Delta S$ & Entropy of fusion & 130 j kg$^{-1}$ K$^{-1}$& C \\
    
    \hline\multicolumn{4}{c}{Upper mantle}\\\hline
    
    -- & Material  & olivine & A \\
    $\rho_0$, $K_0$, $K_0'$, $\gamma_0$, $q$, $\theta_0$ & Equation of state parameters & 3347 kg m$^{-3}$, 126.8 GPa, 4.274, 0.99, 2.1, 809 K & A \\ 
    $B$, $n$, $E^*$, $\dot \epsilon$ & Viscosity parameters & $3.5\times 10 ^{-15}$ Pa$^{-n}$s$^{-1}$, 3, 430$\times 10^{3}$ j mol$^{-1}$, $1\times 10 ^{-15}$ s$^{-1}$ & D \\
    $V^*$ & Activation volume & 2.5 $\times 10^{-6}$ m$^3$ mol$^{-1}$ & F \\
    $\alpha$ & Thermal expansivity & 3.6 $\times 10^{-6}$ K$^{-1}$ & D \\
    $c_p$ & Specific heat  & 1250 j kg$^{-1}$ K$^{-1}$& C \\
    $k_m$ & Thermal conductivity & 6 W m$^{-1}$ K$^{-1}$ & C \\
    $\kappa_m$ & Thermal diffusivity & $7.5\times 10^{-7}$ m$^2$ s$^{-1}$ & E  \\
    $\theta$ & Potential temperature & 1700 K  & F \\
    $\chi_r$ & Radiactive heat correction & 1.253 & -- \\ 
    
    \hline
  \end{tabular}
  \caption{Reference Thermal Evolution Model parameters. Sources: (A)
    VAL06, (B) \citet{Nimmo09a}, (C) \citet{Gaidos10}, (D)
    \citet{Tachinami11}, (E) \citet{Ricard09}, (F)
    \citet{Stamenkovic11}.
    \label{tab:ThermalModelParameters}}
    %REVISED-MAN
\end{table*}
%TTTTTTTTTTTTTTTTTTTTTTTTTTTTTTTTTTTTTTTTTTTTTTTTTTTTTTTTTTTTTTTTTTTTTTTTTTT

%============================================================
\section{Planetary magnetic field}
\label{sec:ScalingPMF}
%============================================================

In recent years improved numerical experiments have constrained the
full set of possible scaling laws used to predict the properties of
planetary and stellar convection-driven dynamos (see
\citealt{Christensen10} and references therein).  It has been found
that in a wide range of physical conditions the global properties of a
planetary dynamo can be expressed in terms of simple power-law
functions of the total convective power $Q_\rom{conv}$ and the size of
the convective region.
%REVISED-MAN

One of the most important results of power-based scaling laws is the
fact that the volume averaged magnetic field intensity
$B_\rom{rms}^2=(1/V)\int B^2 dV$ does not depend on the rotation rate
of the planet (eq. (6) in \citealt{Zuluaga12}),
%REVISED-MAN

\beq{eq:Brms_scaling}
B_\rom{rms}\approx C_\rom{Brms}\; \mu_0^{1/2} {\bar{\rho}_c}^{1/6} (D/V)^{1/3}
Q_\rom{conv}^{1/3}
\eeq

Here $C_\rom{Brms}$ is a fitting constant obtained from numerical
dynamo experiments and its value is different in the case of dipolar
dominated dynamos, $C^{\rm dip}_\rom{Brms}=0.24$, and multipolar dynamos,
$C^{\rm mul}_\rom{Brms}=0.18$. $\bar{\rho}_c$, $D=R_\star-R_\rom{ic}$ and
$V=4\pi (R_\star^3-R_\rom{ic}^3)/3$ are the average density, height
and volume of the convective shell.
%REVISED-MAN

%% We
%% are assuming that the whole external liquid iron core is convecting.
%% In a real case only a fraction of the core volume is involved in
%% dynamo action and therefore the magnetic fields predicted with
%% equation \ref{eq:Brms_scaling} and with our assumption will
%% underestimate the actual field strenght \citep{Gaidos11}.  We have,
%% however, verified that this effect is only important for time periods
%% much longer than the time it takes to start the inner core nucleation.
%% For a planet with the same mass as the Earth the time during which the
%% stellar aggression is the largest is much less than that time for
%% inner core nucleation (see section \ref{sec:Results}).
%REVISED2

The dipolar field intensity at the planetary surface, and hence the
dipole moment of the PMF, can be estimated if we have information
about the power spectrum of the magnetic field at the core surface.
Although we cannot predict the relative contribution of each mode to
the total core field strength, numerical dynamos exhibit an
interesting property: there is a scalable dimensionless quantity, the
local Rossby number $Ro^*_l$, that could be used to distinguish
dipolar dominated from multipolar dynamos.  The scaling relation for
$Ro^*_l$ is (eq. (5) in \citealt{Zuluaga12}):
%REVISED-MAN

%% We
%% are assuming that the whole external liquid iron core is convecting.
%% In a real case only a fraction of the core volume is involved in
%% dynamo action and therefore the magnetic fields predicted with
%% equation \ref{eq:Brms_scaling} and with our assumption will
%% underestimate the actual field strenght \citep{Gaidos11}.  We have,
%% however, verified that this effect is only important for time periods
%% much longer than the time it takes to start the inner core nucleation.
%% For a planet with the same mass as the Earth the time during which the
%% stellar aggression is the largest is much less than that time for
%% inner core nucleation (see section \ref{sec:Results}).
%REVISED2

\beq{eq:Roml_scaling}
Ro^*_l = C_\rom{Rol}\; {\bar{\rho}_c}^{-1/6} R_c^{-2/3} D^{-1/3} V^{-1/2}
Q_\rom{conv}^{1/2} P^{7/6}.
\eeq

Here $C_\rom{Rol}=0.67$ is a fitting constant and $P$ is the period of
rotation.  It has been found that dipolar dominated fields arise
systematically when dynamos have $Ro^*_l<0.1$.  Multipolar fields
arise in dynamos with values of the local Rossby number close to and
larger than this critical value.  From eq. (\ref{eq:Roml_scaling}) we
see that in general fast rotating dynamos (low $P$) have dipolar
dominated core fields while slowly rotating ones (large $P$)
produce multipolar fields and hence fields with a much lower dipole
moment.
%REVISED-MAN

It is important to stress that the almost independence of
$B_\rom{rms}$ on rotation rate, together with the role that rotation
has in the determination of the core field regime, implies that even
very slowly rotating planets could have a magnetic energy budget of
comparable sized than rapidly rotating planets with similar size and
thermal histories.  In the former case the magnetic energy will be
redistributed among other multipolar modes rendering the core field
more complex in space and probably also in time.  Together all these
facts introduce a non-trivial dependence of dipole moment on rotation
rate very different from that obtained with the traditional scaling
laws used in previous works (see e.g. \citealt{Griebmeier04} and
\citealt{Khodachenko07}).  Here we want to emphasize a property that
was also previously overlooked.  Multipolar dominated dynamos produces
magnetic fields that decays more rapidly with distance than dipolar
fields and so it is expected that a planet with a multipolar magnetic
field will be less protected than those having strongly dipolar
fields.
%REVISED-MAN

Using the value of $B_\rom{rms}$ and $Ro^*_l$ we can compute the {\it
  maximum dipolar component} of the field at core surface.  For this
purpose we use an upper bound to the dipolarity fraction $f_\rom{dip}$
(the ratio of the dipolar component to the total field strength at
core surface).  Dipolar dominated dynamos have by definition
$f_\rom{dip}\leq f_\rom{dip}^{max}=1.0$.  The case of reversing
dipolar and multipolar dynamos is more complex.  Numerical dynamo
experiments show that multipolar dynamos have $Ro^*_l\gtrsim 0.1$ and
$f_\rom{dip}\lesssim f_\rom{dip}^{max}=0.35$.  However to avoid
inhomogeneities in the transition region around $Ro^*_l\approx 0.1$ we
calculate a maximum dipolarity fraction through a ``soft step
function'',
$f_\rom{dip}^{max}=\alpha+\beta/\{\exp[(\Rol-0.1)/\delta]+1\}$ with
$\alpha$, $\beta$ and $\delta$ numerical constants that fits the
envelop of the numerical dynamo data (see upper panel of figure 1 in
\citealt{Zuluaga12}).

To connect this ratio to the volumetric averaged magnetic field
$B_\rom{rms}$ we use the volumetric dipolarity fraction $b_\rom{dip}$
that it is found, as shown by numerical experiments, conveniently
related with the maxium value of $f_\rom{dip}$ through eq. (12) in
\citealt{Zuluaga12},
%REVISED-MAN

\beq{eq:bdip_fdip}
b_\rom{dip}^{min}=c_\rom{bdip} {f_\rom{dip}^{max}}^{-11/10}
\eeq

where $c_\rom{bdip}\approx 2.5$ is again a fitting constant.  It is
important to notice here that the exponent $11/10$ is the ratio of the
smallest integers close to the numerical value of the fitting exponent
(see figure 1 in \citealt{Zuluaga12}).  We use this convention
following \citealt{Olson06}.

Finally by combining eqs. (\ref{eq:Brms_scaling})-(\ref{eq:bdip_fdip})
we can compute an upper bound to the dipolar component of the field at
the CMB:
%REVISED-MAN

\beq{eq:Bdipmax_scaling}
B_{c}^{dip}\lesssim \frac{1}{b_\rom{dip}^{min}} B_\rom{rms} = 
\frac{{f_\rom{dip}^{max}}^{11/10}}{c_\rom{bdip}}\;B_\rom{rms}
\eeq

The surface dipolar field strength is estimated using,

\beq{eq:Bdip_surface}
B_p^{dip}(R_p)=B_c^{dip}\left(\frac{R_p}{R_c}\right)^3
\eeq

and finally the total dipole moment is calculated using
eq. (\ref{eq:M}) for $r=R_p$.

It should be emphasized that the surface magnetic field intensity
determined using eq.  (\ref{eq:Bdip_surface}) overestimates the PMF
dipolar component.  The actual field could be much more complex
spatially and the dipolar component could be lower.  As a consequence
our model can only predict the maximum level of protection that a
given planet could have from a dynamo-generated intrinsic PMF.
%REVISED-MAN

\begin{figure}  
  \centering
   \includegraphics[width=0.7
     \textwidth]{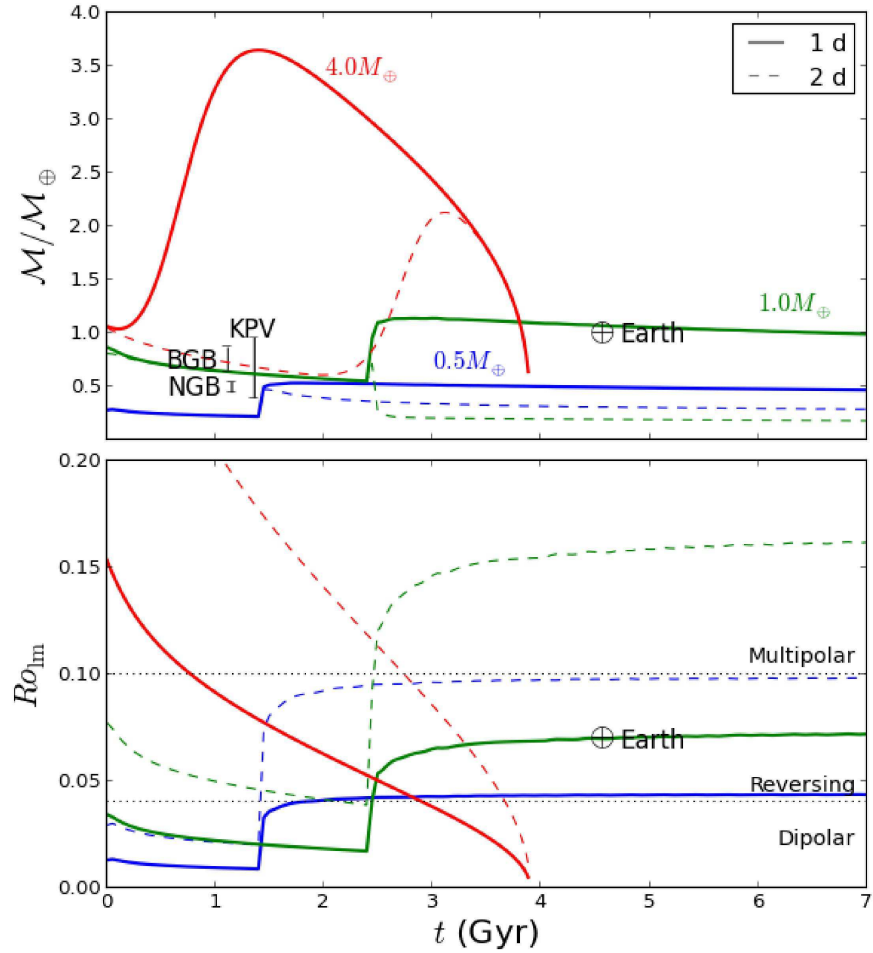}
  \scriptsize
  \caption{PMF properties predicted using the
    RTEM and eqs. (\ref{eq:Brms_scaling}),
    (\ref{eq:Roml_scaling}) and (\ref{eq:Bdipmax_scaling}) for TPs 0.5-4.0 $\ME$.  
    We plot the local Rossby number
    (lower panel), the maximum surface dipole field (middle panel) and
    the maximum dipole moment (upper panel).  We included the present values 
    of the geodynamo ($\oplus$ symbol) and three measurements of paleomagnetic
    intensities (error bars) at 3.2 and 3.4 Gyr ago \citep{Tarduno10}.
    We compare the magnetic properties for two periods of rotation, 1
    day (solid curves) and 2 days (dashed curves).  The effect of a
    larger period of rotation is more significant at early times in
    the case of massive planets ($M_p\gtrsim 2\ME$) and at late times
    for lower mass planets.\vspace{0.2cm}} \label{fig:MagneticField}
    %REVISED-MAN
\end{figure}

\begin{figure}  
  \centering
   \includegraphics[width=0.6
   \textwidth]{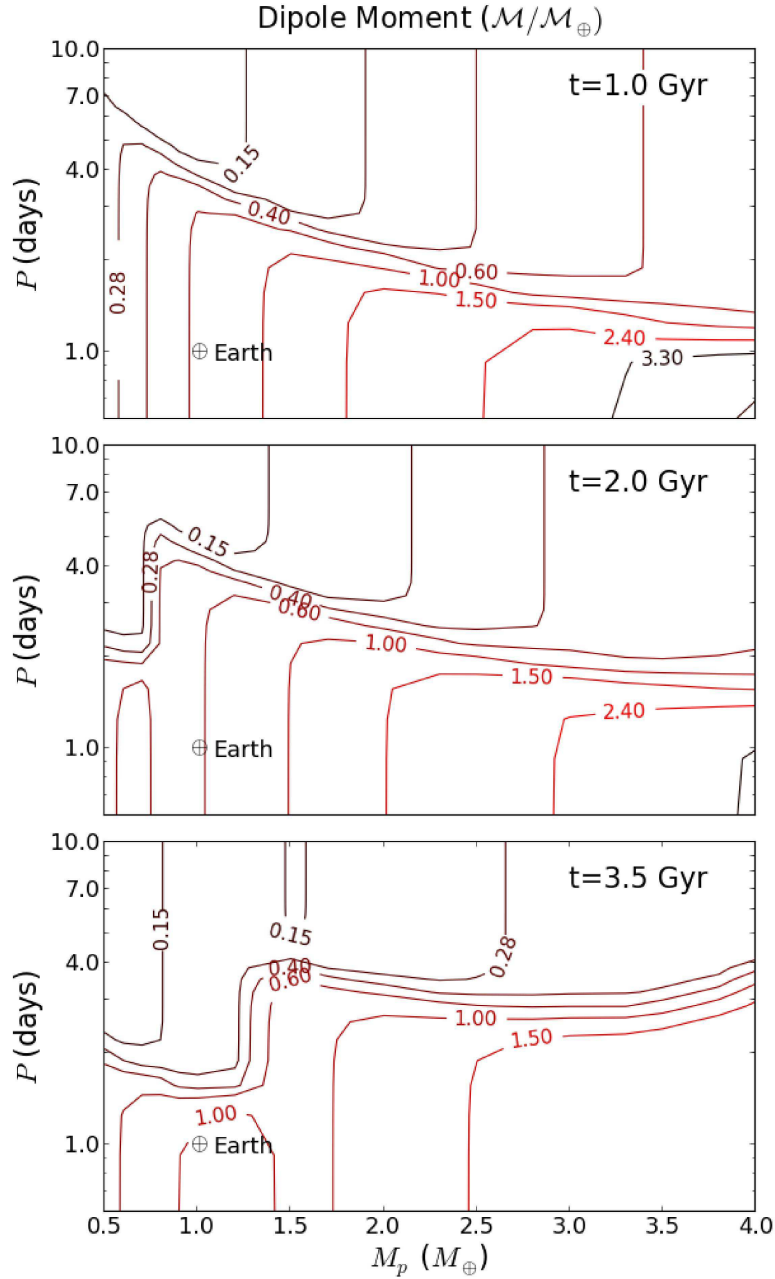}
  \scriptsize 
  \caption{Mass-Period (M-P) diagrams of the dipole moment for
    long-lived planetary dynamos using the RTEM.  Three
    regimes are identified \citep{Zuluaga12}: rapid rotating planets
    ($P\ls1$ day), dipole moments are large and almost
    independent of rotation rate; slowly rotating planets
    ($1\,\rom{day}\ls P \ls 5\,\rom{day}$), dipole moments are
    intermediate in value and highly dependent on rotation rate; and
    very slowly rotating planets ($P\gs 5-10\,\rom{days}$), small
    but non-negligible rotation-independent dipole moments.  For 
    ($M_p<2\ME$) the shape of the dipole-moment contours is
    determined by $t_{ic}$.
    \label{fig:DipoleMoment}}
    %REVISED-MAN
\end{figure}

The results of applying the RTEM to calculate the properties of the
magnetic field of TPs in the mass range 0.5-4.0 $\ME$ using the
scaling laws in eqs. (\ref{eq:Brms_scaling}), (\ref{eq:Roml_scaling})
and (\ref{eq:Bdipmax_scaling}) are summarized in figures
\ref{fig:MagneticField} and \ref{fig:DipoleMoment}.  In figure
\ref{fig:MagneticField} we show the local Rossby number, the maximum
dipolar field intensity and the dipole moment as a function of time
computed for planets with different mass and two different periods of
rotation ($P=1$ day and $P=2$ days).  This figure shows the effect that
rotation has on the evolution of dynamo geometry and hence in the
maximum attainable dipolar field intensity at the planetary surface.
In figure \ref{fig:DipoleMoment} we have summarized in mass-period
(M-P) diagrams \citep{Zuluaga12} the evolution of the dipole moment
for planets with long-lived dynamos.  We see that for periods lower
than 1 day and larger than 5-7 days the dipole moment is nearly
independent of rotation.  Slowly rotating planets have a
non-negligible dipole moment which is systematically larger for more
massive planets.
%REVISED-MAN

%============================================================
\section{Planet-Star interaction}
\label{sec:PlanetStarInteraction}
%============================================================

The PMF properties constrained using the thermal evolution model and
the dynamo-scaling laws are not enough to evaluate the level of
magnetic protection of a potentially habitable TP.  We also need to
estimate also the magnetosphere and stellar properties (stellar wind
and luminosity) as a function of time in order to assess properly the
level of star-planet interaction.
%REVISED-MAN

Since the model developed in previous sections provides only the
maximum intensity of the PMF, we will be interested here into
constraint the magnetopshere and stellar properties from below,
i.e. to find the lower level of ``stellar aggression'' for a given
star-planet configuration.  Combining upper bounds of PMF properties
and lower bounds for the star-planet interaction will produce an
overestimation of the overall magnetic protection of a planet.  If
under this model a given star-planet configuration is not suitable to
provide enough magnetic protection to the planet, the actual case
should be much worse.  But if, on the other hand, our upper-limit
approach predicts a high level of magnetic protection, the actual case
could still be that of an unprotected planet.  Therefore our model is
capable at predicting which planets will be unprotected but less able
when predicting which ones will be actually protected.
%REVISED-MAN

%......................................................................
\subsection{The Habitable Zone (HZ) and tidally locking limits}
\label{subsec:HZTL}
%......................................................................

Surface temperature and hence ``first-order'' habitability of a planet
depends on three basic factors: 1) the fundamental properties of the
star (luminosity $L_\star$, effective temperature $T_\rom{\star}$ and
radius $R_\star$) 2) the average star-planet distance (distance to the
HZ) and 3) the conmensurability of planetary rotation and orbital
period (tidal locking).  These properties should be properly modelled
in order to assess the degree of star-planet interaction which are
critical at determining the magnetic protection.
%REVISED-MAN

The basic properties of main-sequence stars of different masses and
metallicities have been studied for decades and are becoming critical
in assessing the actual properties of newly discovered exoplanets.
The case of low mass main sequence stars (GKM) are particularly
important in providing the properties of the stars with the highest
potential to harbor habitable planets with evolved and diverse
biospheres.
%REVISED-MAN

In this work we will use the theoretical results by
\citealt{Baraffe98} (hereafter BAR98) that predict the evolution of
different metallicities main-sequence GKM stars.  We have chosen from
that model those results corresponding to the case of solar
metallicity stars.  We have disregarded the fact that the basic
stellar properties actually evolve during the critical period where
magnetic protection will be evaluated, i.e. $t=0.5-3$ Gyr.  To be
consistent with the purpose of estimating upper limits to magnetic
protection, we took the stellar properties as provided by the model at
the highest end of the time interval, i.e. $t=3$ Gyr.  Since
luminosity increases with time in GKM stars this assumption guarantees
the largest distance of the HZ and hence the lowest effects of the
stellar insolation and the stellar wind.
%REVISED-MAN

\begin{figure}  
  \centering
   \includegraphics[width=0.8
     \textwidth]{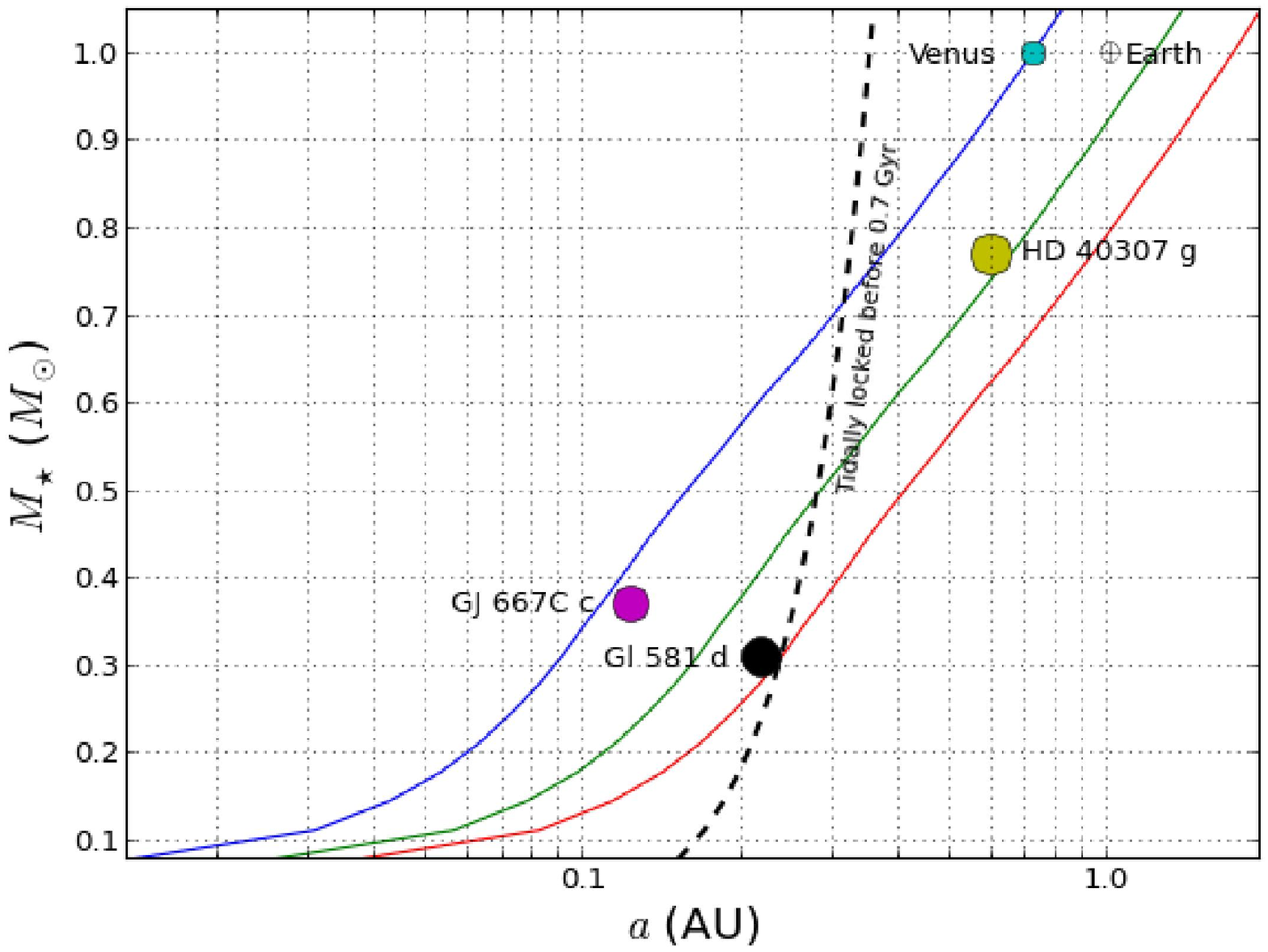}
  \caption{HZ limits corresponding to the conservative criteria of
    recent Venus and early Mars according to the updated limits
    estimated by \citet{Kopparapu13}.  Stellar properties are computed
    at $\tau=3$ Gyr using the models by \citealt{Baraffe98}. Planets
    at distances below the dashed line would be tidally locked before
    0.7 Gyr \citep{Peale77}.  The location of Earth, Venus and the
    potentially habitable extra-solar-system planets GL 581d, GJ 667 C
    c and HD 40307g are also
    included.\vspace{0.2cm}} \label{fig:StellarProperties}
    %REVISED-MAN
\end{figure}

In order to estimate the HZ limits we use the recently updated values
calculated by \citet{Kopparapu13}.  In particular we use the
interpolation formula in eq. (2) and coefficients in Table (2) to
compute the most conservative limits of Recent Venus and Early Mars.
The limits calculated for the stellar properties assumed here are
depicted in figure \ref{fig:StellarProperties}.

%% Well-known results of \citealt{Kasting93} and the the parabolic
%% fitting developed by \citealt{Selsis07a}.  Accordingly the inner
%% and outer limits of the HZ, $a_\rom{in}$ and $a_\rom{out}$
%% respectively, are given in terms of the stellar effective
%% temperature $T_*$ and luminosity $L$ by: %REVISED-MAN

%% \beq{eq:HZ}
%% \begin{array}{lll}
%% a_\rom{in} & = & ( a_{in\odot}-\alpha_\rom{in} \Delta T -
%% \beta_\rom{in} \Delta T^2 ) \sqrt{L/L_\odot} \\
%% a_\rom{out} & = & ( a_{out\odot}-\alpha_\rom{out} \Delta T -
%% \beta_\rom{out} \Delta T^2 ) \sqrt{L/L_\odot}
%% \end{array}
%% \eeq

%% where $a_{in\odot}$ and $a_{out\odot}$ are the inner and outer limits
%% of the HZ for the present sun, $\Delta T=T_*-T_\odot$ and $\alpha$ and
%% $\beta$ are fitting constants.  For our purposes we use the
%% conservative limits given by the criteria of ``recent Venus'' and
%% ``early Mars'' \citep{Kasting93}.  For this case $a_{in\odot}=0.72$,
%% $\alpha_\rom{in}=2.7619\times 10^{-5}$, $\beta_\rom{in}=3.8095\times
%% 10^{-9}$ and $a_{out\odot}=1.77$, $\alpha_\rom{out}=1.3786\times
%% 10^{-4}$, $\beta_\rom{out}=1.4286\times 10^{-9}$ \citep{Selsis07a}.
%% %REVISED-MAN

The orbital and rotational properties of planets at close-in orbits
are strongly affected by the gravitational and tidal interaction with
the host star.  Tidal torque dampens the primordial rotation and axis
tilt leaving the planet in a final resonant equilibrium where the
period of rotation $P$ becomes commensurable with the orbital period
$P_o$,
%REVISED-MAN

\beq{eq:P-resonance}
P:P_o=n:2
\eeq

Here $n$ is an integer larger than or equal to 2.  The value of $n$ is
determined by multiple dynamic factors, the most important being the
orbital planetary eccentricity
\citep{Leconte10,FerrazMello08,Heller11}.  In the solar system the
tidal interaction between the Sun and Mercury has trapped the planet
in a 3:2 resonance.  In the case of GL 581d, detailed dynamic models
predict a resonant 2:1 equilibrium state \citep{Heller11}, i.e. the
rotation period of the planet is a half of its orbital period.
%REVISED-MAN

Although estimating in general the time required for the ``tidal
erosion'' is very hard given the large uncertainties in the key
physical parameters involved (see \citealt{Heller11} for a detailed
discussion) the maximum distance $a_\rom{tid}$ at which a solid planet
in a circular orbit becomes tidally locked before a given time $t$ can
be roughly estimated by \citep{Peale77}:
%REVISED-MAN

\beq{eq:TidalLimit}
a_\rom{tid}(t) =
0.5\,\rm{AU}\,\left[\frac{(M_\star/M_\odot)^2P_\rom{prim}}{Q}\right]^{1/6}
t^{1/6}
\eeq

Here the primordial period of rotation $P_\rom{prim}$ should be
expressed in hours, $t$ in Gyr and $Q$ is the dimensionless
dissipation function.  For the purposes of this work we assume a
primordial period of rotation $P_\rom{prim} = 17\,\rm{hours}$
\citep{Varga98,Denis11} and a dissipation function $Q \approx 100$
\citep{Henning09, Heller11}.
%REVISED-MAN

In figure \ref{fig:StellarProperties} we summarize the properties of
solar metallicity GKM main sequence stars provided by the BAR98 model
and the corresponding limits of the HZ and tidal locking maximum
distance.  The properties of the host stars of the already discovered
potentially habitable super-Earths, GL 581d, GJ 667Cc and HD 40307g,
are also highlighted in this figure.
%REVISED-MAN

%......................................................................
\subsection{Stellar wind}
\label{subsec:StellarWind}
%......................................................................

The stellar wind and cosmic rays pose the highest risks for a
magnetically unprotected potentially habitable terrestrial planet. The
dynamic pressure of the wind is able to obliterate an exposed
atmosphere, especially druing the early phase of stellar evolution
\citep{Lammer03}, and energetic stellar cosmic rays could pose a
serious risk to any form of surface life directly exposed to them
\citep{Griebmeier05}.
%REVISED-MAN

The last step in order to estimate the magnetospheric properties and
hence the level of magnetic protection is predicting the stellar wind
properties for different stellar masses and as a function of planetary
distance and time.
%REVISED-MAN

There are two simple models used to describe the spatial structure and
dynamics of the stellar wind: the pure hydrodynamical model developed
originally by \citealt{Parker58} that describes the wind as a
non-magnetized, isothermal and axially symmetric flux of particles
(hereafter the {\it Parker's model}) and the more detailed albeit
simpler magneto-hydrodynamic model developed originally by
\citealt{Weber67} that takes into account the effects of stellar
rotation and treats the wind as a magnetized plasma.
%REVISED-MAN

It has been shown that Parker's model describes reliably the
properties of the stellar winds in the case of stars with periods of
rotation of the same order of the present solar value, i.e. $P\sim 30$
days \citep{Preusse05}.  However for rapidly rotating stars,
i.e. young stars and/or active dM stars, the isothermal model
underestimates the stellar wind properties almost by a factor of 2
\citep{Preusse05}. For the purposes of scaling the properties of the
planetary magnetospheres, (equations
(\ref{eq:BmpPsw})-(\ref{eq:Apc})), an underestimation of the stellar
wind dynamic pressure of that size, will give us values of the key
magnetospheric properties that will be off by 10-40\% of the values
given by more detailed models.  Magnetopause fields that have the
largest uncertainties will be underestimated by $\sim 40\%$, while
standoff distances and polar cap areas will be respectively under and
overestimated by just $\sim 10\%$.
%REVISED-MAN

According to Parker's model the stellar wind average particle
velocity $v$ at distance $d$ from the host star is obtained by solving
the {\it Parker's wind} equation \citep{Parker58}:
%REVISED-MAN

\beq{eq:ParkerEquation}
u^2-\log u=4 \log \rho + \frac{4}{\rho} - 3
\eeq

where $u=v/v_c$ and $\rho=d/d_c$ are the velocity and distance
normalized with respect to $v_c=\sqrt{k_B T/m}$ and $d_c=G\Ms m/(4 k_B
T)$ which are respectively the local sound velocity and the critical
distance where the stellar wind becomes subsonic.  $T$ is the
temperature of the plasma which in the isothermal case is assumed
constant at all distances and equal to the temperature of the stellar
corona.  $T$ is the only free parameter controlling the velocity
profile of the stellar wind.
%REVISED-MAN

The number density $n(d)$ is calculated from the velocity using the
continuity equation:
%REVISED-MAN

\beq{eq:Parker_n}
n(d)=\frac{\dot{\Ms}}{4\pi d^2 v(d) m}
\eeq

Here $\dot{\Ms}$ is the stellar mass-loss rate, which is a free
parameter in the model.  
%REVISED-MAN

To calculate the evolution of the stellar wind we need a way to
estimate the evolution of the coronal temperature $T$ and the
mass-loss rate $\dot{\Ms}$.
%REVISED-MAN

Using observational estimates of the stellar mass-loss rate
\citep{Wood02} and theoretical models for the evolution of the stellar
wind velocity \citep{Newkirk80}, \citealt{Griebmeier04} and
\citealt{Lammer04} developed semiempirical formulae to calculate the
evolution of the long-term averaged number density and velocity of the
stellar wind for main sequence stars at a given reference distance (1
AU):
%REVISED-MAN

\beq{eq:vt}
v_{1\rm{AU}}(t)=v_0 \left(1+\frac{t}{\tau}\right)^{\alpha_v}
\eeq

\beq{eq:nt}
n_{1\rm{AU}}(t)=n_0 \left(1+\frac{t}{\tau}\right)^{\alpha_n}
\eeq

Here $\alpha_v=-0.43$, $\alpha_n=-1.86\pm 0.6$ and
$\tau=25.6\,\rm{Myr}$ \citep{Griebmeier09}. The parameters $v_0=3971$
km/s and $n_0=1.04\times 10^{11}\,\rm{m}^{-3}$ are estimated from the
present long-term averages of the solar wind as measured at the
distance of the Earth $n(4.6\,\rm{Gyr},1\,\rm{AU},1\,\MSUN)=6.59\times
10^6\,\rm{m}^{-3}$ and
$v(4.6\,\rm{Gyr},1\,\rm{AU},1\,\MSUN)=425\,\rm{km/s}$
\citep{Schwenn90}.
%REVISED-MAN

Using these formulae \citealt{Griebmeier07a} devised a clever way to
estimate consistently $T(t)$ and $\dot{\Ms}(t)$ in the Parker's model
and hence we are able predict the stellar wind properties as a
function of $d$ and $t$.  For the sake of completeness we summarize
here this procedure.  For further details see section 2.4 in
\citealt{Griebmeier07a}
%REVISED-MAN

For a stellar mass $\Ms$ and time $t$, the velocity of the stellar
wind at $d=1$ AU, $v_{1\rm{AU}}$, is calculated using equation
(\ref{eq:vt}).  Replacing this velocity in the Parker's wind equation
for $d=1$ AU, we find numerically the temperature of the Corona
$T(t)$.  This parameter is enough to provide us the whole velocity
profile $v(t,d,\Ms)$ at time $t$.  To compute the number density we
need the mass-loss rate for this particular star and at this time.
Using the velocity and number density calculated from
eqs. (\ref{eq:vt}) and (\ref{eq:nt}) the mass-loss rate for the Sun
$\dot{M_\odot}$ at time t and $d=1$ AU can be obtained:
%REVISED-MAN

\beq{eq:dotMsun}
\dot M_\odot(t)=4\pi (1\,\rm AU)^2\;m\;
n_{1\rm{AU}}(t)
v_{1\rm{AU}}(t)
\eeq

Assuming that the mass-loss rate scales-up simply with the stellar
surface area, i.e. $\dot M_\star(t)=\dot
M_\odot(t)(R_\star/R_\odot)^2$, the value of $\dot{\Ms}$ can finally
be estimated.  Using $v(t,d,\Ms)$ and $\dot{\Ms}$ in the continuity
equation (\ref{eq:Parker_n}), the number density of the stellar wind
$n(t,d,\Ms)$ is finally obtained.
%REVISED-MAN

\begin{figure}
  \centering
  \includegraphics[width=0.8
  \textwidth]{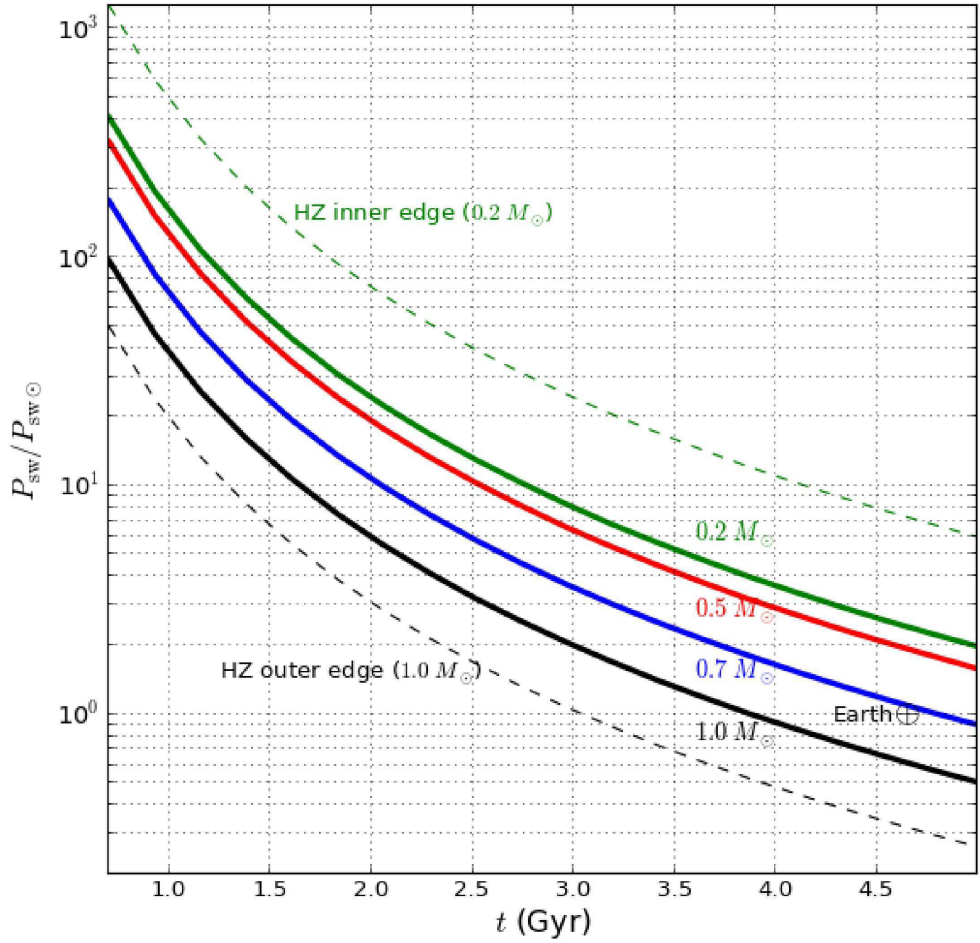}
  \caption{Evolution of the stellar wind dynamic pressure at the
    center of the HZ for a selected set of stellar masses.  The
    reference average solar wind pressure is $P_{\rm{SW}\odot}=1.86$
    nPa.  Dashed curves indicate the value of the stellar wind
    pressure at the inner and outer edges of the HZ around stars with
    0.2 $M_{\odot}$ and 1.0 $M_{\odot}$ respectively. The HZ limits
    where the pressure were calculated are assumed static and equal to
    those at $\tau=3$ Gyr.  Stellar wind pressure at $t<0.7$ Gyr
    computed with the semiempirical model used in this work is too
    uncertain and were not plotted.\vspace{0.5cm}}
  \label{fig:SW}
  %REVISED-MAN
\end{figure}

The value of the stellar wind dynamic pressure
$P_\rom{dyn}(t,d,\Ms)=m\,n(t,d,\Ms)\,v(t,d,\Ms)^2$ inside the HZ of
four different stars as computed using the procedure described before
is plotted in figure \ref{fig:SW}.
%REVISED-MAN

It is important to stress here that for stellar ages $t\ls 0.7$ Gyr
the semiempirical formulae in eqs. (\ref{eq:vt}) and (\ref{eq:nt}) are
not longer reliable \citep{Griebmeier07a}.  These equations are based
in the empirical relationship observed between the X-ray surface flux
and the mass-loss rate $\dot{\Ms}$ \citep{Wood02,Wood05} which has
been reliably obtained only for ages $t\gs 0.7$ Gyr.  However
\citealt{Wood05} have shown that an extrapolation of the empirical
relationship to earlier times overestimates the mass-loss rate by a
factor of 10-100.  At times $t\ls 0.7$ Gyr and over a given magnetic
activity threshold the stellar wind of main-sequence stars seems to be
inhibited \citep{Wood05}.  Therefore the limit placed by observations
at $t\approx 0.7$ Gyr is not simply an observational constraint but
could mark the time where the early stellar wind also reaches a
maximum (J.L. Linsky, Private Communication).  This fact suggests that
at early times the effect of the stellar wind on the planetary
magnetosphere is much lower than normally assumed.  Hereafter we will
assume that intrinsic PMF are strong enough to protect the planet at
least until the maximum of the stellar wind is reached at $t\approx
0.7$ Gyr and focus on the stellar-wind and magnetosphere properties
for times larger than this.
%REVISED-FIN

%%%%%%%%%%%%%%%%%%%%%%%%%%%%%%%%%%%%%%%%%%%%%%%%%%%%%%%%%%%%%%%%%%%%%%%%%
\section{Results}
\label{sec:Results}
%%%%%%%%%%%%%%%%%%%%%%%%%%%%%%%%%%%%%%%%%%%%%%%%%%%%%%%%%%%%%%%%%%%%%%%%%

Using the results of our RTEM, the power-based scaling laws for dynamo
properties, and the properties of the stellar insolation and stellar
wind pressure, we have calculated the magnetosphere properties of
earth-like planets and super-Earths in the HZ of different
main-sequence stars.  We have performed these calculations for
hypothetical TPs in the mass-range 0.5-6 $\ME$ and for the already
discovered potentially habitable planets GL 581d, GJ 667Cc and HD
40307g (see table \ref{tab:SEs}).  The case of the Earth and an
habitable Venus has also been studied for references purposes.
%REVISED-MAN

%TTTTTTTTTTTTTTTTTTTTTTTTTTTTTTTTTTTTTTTTTTTTTTTTTTTTTTTTTTTTTTTTTTTTTT
\begin{table*}[ht]
  \centering
  \scriptsize
  \begin{tabular}{cccccccccccc}
    \hline\hline 
    Planet & $M_p(\ME)$ & $R_p(R_{\oplus})$ & a(AU) & $P_o$ (days) & e &
    S-type & $M_\star(M_{\odot})$ & age(Gyr) & tid.locked & Refs.\\ \hline\hline
     Earth & 1.0 & 1.0 & 1.0 & 365.25 & 0.016 & G2V & 1.0 & 4.56 & No & --\\
     Venus & 0.814 & 0.949 & 0.723 & 224.7 & 0.007 & G2V & 1.0 & 4.56 & Probably & --\\
     GJ 667Cc & 4.545 & 1.5* & 0.123 & 28.155 & $<0.27$ & M1.25V & 0.37 & $>2.0$ & Yes & (1)\\
%    HD 85512b & 3.496 & 1.4* & 0.26 & 58.43 & 0.11 & K5V & 0.69 & $5.6-8.0$ & Yes & .\\
     GL 581d & 6.038 & 1.6* & 0.22 & 66.64 & 0.25 & M3V & 0.31 & $4.3-8.0$ & Yes & (2),(3)\\
     HD 40307g & 7.1 & 1.7* & 0.6 & 197.8 & 0.29 & K2.5V & 0.77 & 4.5 & No & (4) \\    
     \hline\hline
  \end{tabular}
  \caption{Properties of the already discovered SEs inside the HZ of
    their host stars. For reference purposes the properties of Venus
    and the Earth are also included.  Values of radii marked with an
    $*$ are unknown and were estimated using the mass-radius relation
    for planets with the same composition as the Earth, i.e. $R_p =
    R_{\oplus} (M_p / \ME)^{0.27}$ (VAL06). References are:
    (1) \citealt{Bonfils11b}, (2) \citealt{Udry07}, (3)
    \citealt{Mayor09}, (4) \citealt{Tuomi2012}
    \label{tab:SEs}}
    %REVISED-MAN
\end{table*}
%TTTTTTTTTTTTTTTTTTTTTTTTTTTTTTTTTTTTTTTTTTTTTTTTTTTTTTTTTTTTTTTTTTTTTT

To include the effect of rotation in the properties of the PMF we have
assumed that planets in the HZ of late K and dM stars ($M<0.7
M_\odot$) are tidally locked at times $t<0.7$ Gyr (n=2 in
eq. (\ref{eq:P-resonance}), see figure \ref{fig:StellarProperties}).
Planets around G and early K stars ($M\gs 0.7 M_\odot$) will be
assumed to have their primordial periods of rotation that we chose in
the range $1-100$ days as predicted by models of planetary formation
\citep{Miguel10}.
%REVISED-MAN

\begin{figure*}[t]
  \centering
   \includegraphics[width=1.0
   \textwidth]{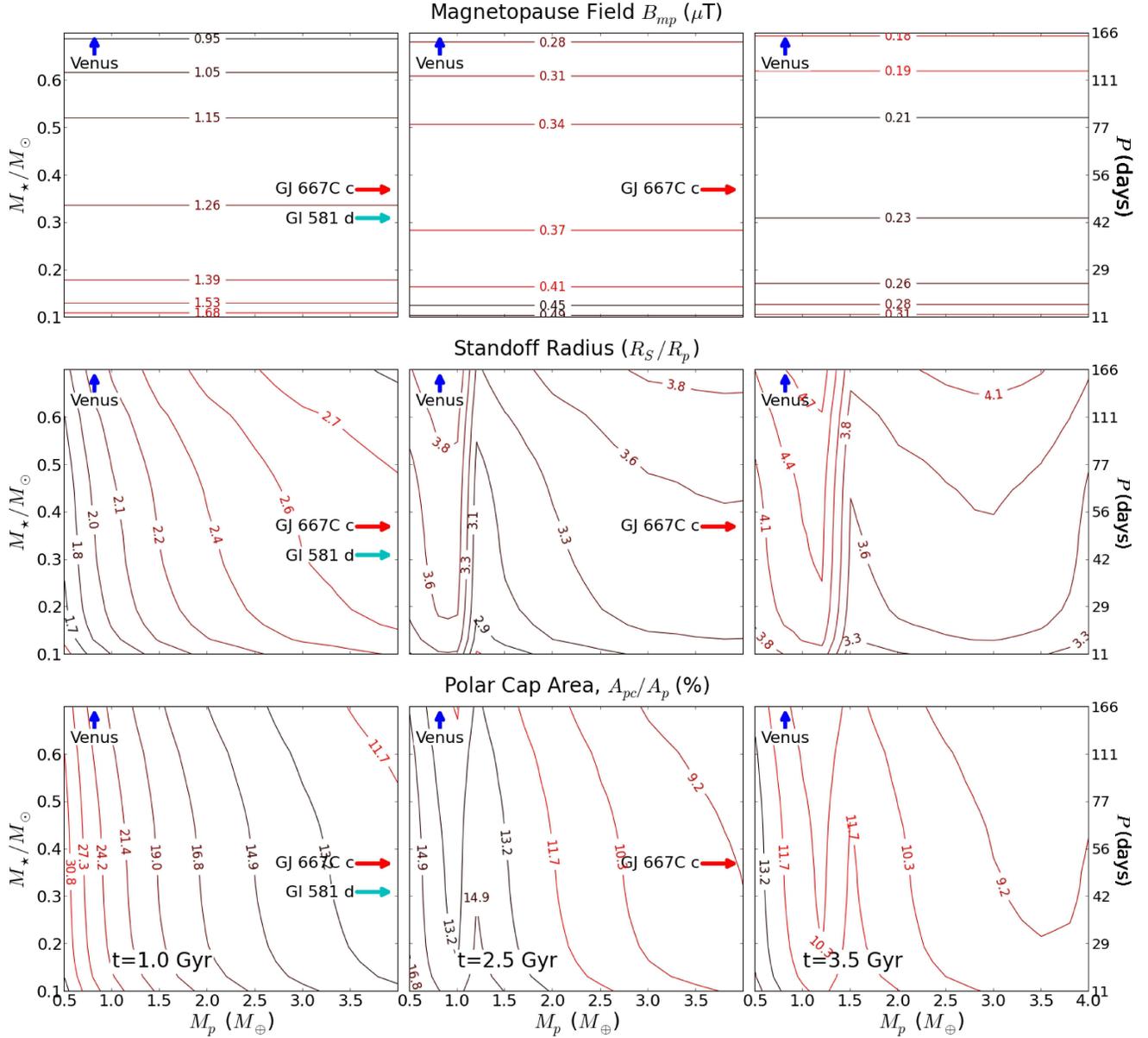}
   \caption{Evolution of the magnetopause field (upper row), standoff
     distance (middle row) and polar cap area (lower row) of tidally
     locked (slow rotating) planets around late dK and dM stars.  The
     rotation of each planet is assumed equal to the orbital
     period at the middle of the HZ (see values in the rightmost
     vertical axes).  The value of the magnetosphere properties
     rerturned by the contour lines in these plots could be an under
     or an overestimation of these properties according to the
     position of a planet inside the HZ.  In the case of GL 581d (GJ
     667Cc), which is located in the outer (inner) edge of the HZ, the
     magnetopause field and polar cap area are overestimated
     (underestimated) while the standoff distance is underestimated
     (overestimated).\vspace{0.2cm}}
  \label{fig:Bmp-Rs-Apc-Tid}
  %REVISED-MAN
\end{figure*}

\begin{figure*}
  \centering
   \includegraphics[width=1.0
   \textwidth]{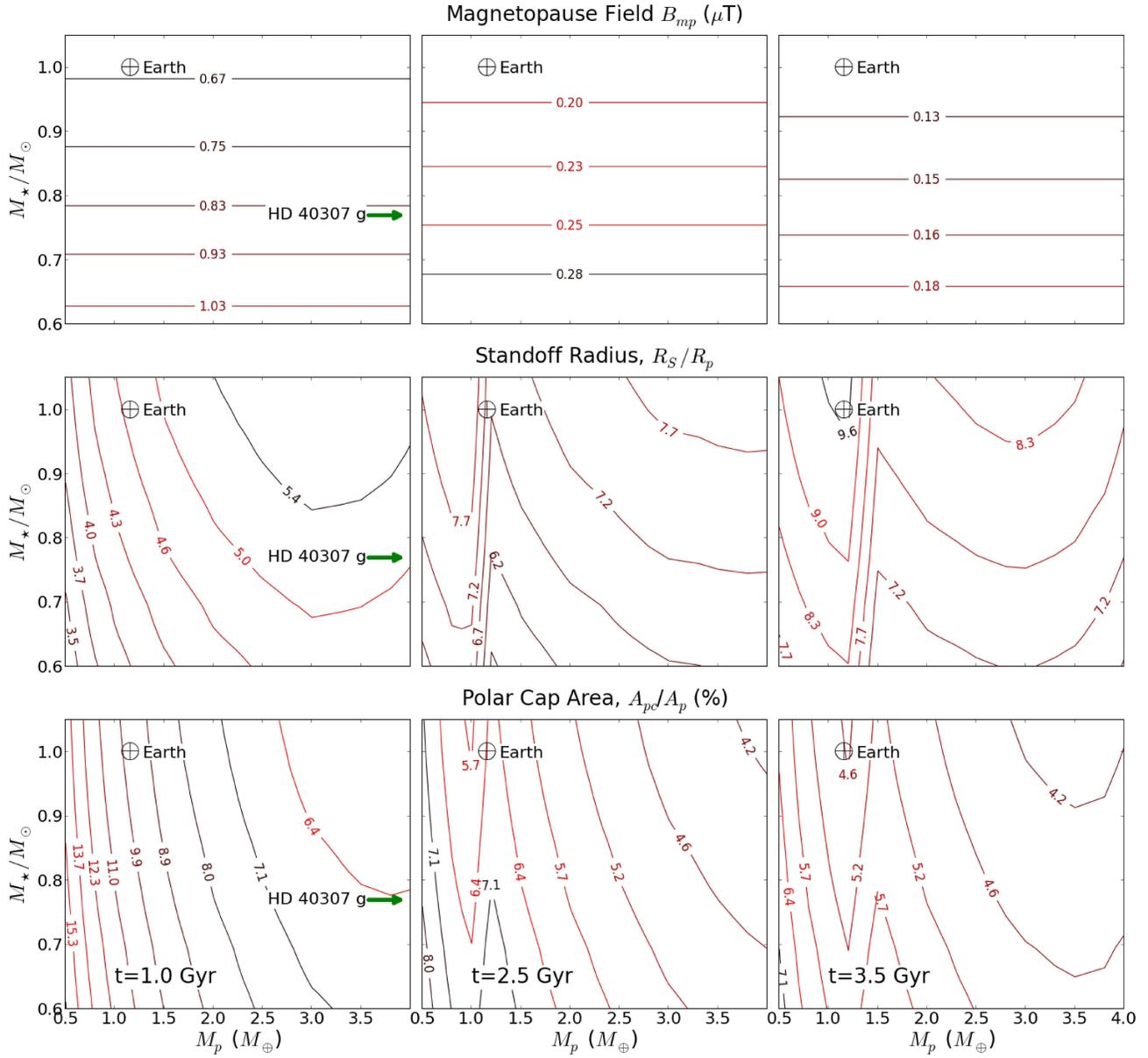}
  \caption{Same as figure \ref{fig:Bmp-Rs-Apc-Tid} but for unlocked
    (fast rotating) planets around early K and G stars ($\Ms\gs 0.7$).
    For all planets we have assumed a constant period of rotation
    $P=1$ day.\vspace{0.5cm}}
  \label{fig:Bmp-Rs-Apc-Notid}
  %REVISED-MAN
\end{figure*}

Figures \ref{fig:Bmp-Rs-Apc-Tid} and \ref{fig:Bmp-Rs-Apc-Notid} show
the evolution of magnetosphere properties for tidally locked and
unlocked potentially habitable planets respectively.  In all cases we
have assumed that the planets are in the middle of the HZ of their
host stars.
%REVISED-MAN

Even at early times tidally locked planets of arbitrary mass have a
non-negligible magnetosphere radius $R_S>1.5 \ R_p$.  Previous
estimates of the standoff distances for tidally locked planets are
much lower than the values reported here.  As an example
\citealt{Khodachenko07} place the standoff distances well below $2
\ R_p$, even under mild stellar wind conditions (see figure 4 in their
work) and independent of planetary mass and age.  In contrast our
model predicts standoff distances for tidally locked planets in the
range of 2-6 $R_p$ depending on planetary mass and stellar age.  The
differences between both predictions arise mainly from the
underestimation of the dipole moment for slowly rotating planets found
in these works.  Thermal evolution and the dependency on planetary
mass of the PMF properties are responsible for the rest of the
discrepancies in previous estimates of the magnetosphere properties.
%REVISED-MAN

Though tidally locked planets seem to have larger magnetospheres than
previously expected, they still have large polar caps, a feature that
was previously overlooked.  As a consequence of this fact well
protected atmospheres, i.e. atmospheres that lie well inside of the
magnetosphere cavity (hereafter {\it magnetised planets}), could have
more than 15\% of their surface area exposed to open field lines where
thermal and non-thermal processes could efficiently remove atmospheric
gases.  Moreover, our model predicts that these planets would have
multipolar PMF
%, hereafter {\it paleomagnetospheres}, 
which contributes to an increase of the atmospheric area open to the
interplanetary and magnetotail regions \citep{Siscoe76}.  Then the
exposition of magnetised planets to harmful external effects would be
a complex function of the standoff distance and the polar cap area.
%REVISED-MAN

Overall magnetic protection improves with time.  As the star evolves
the dynamic pressure of the stellar wind decreases more rapidly than
the dipole moment (see figures \ref{fig:DipoleMoment} and
\ref{fig:SW}).  As a consequence the standoff distance grows in time
and the polar cap is shrunk.  However with the reduction in time of
the stellar wind pressure the magnetopause field is also reduced a
fact that can affect the incoming flux of CR at late times.
%REVISED-MAN

The sinuous shape of the contour lines in the middle and lower rows is
a byproduct of the inner core solidification in planets with
$M_p<2\ME$.  Critical boundaries between regions with very different
behaviors in the magnetosphere properties are observed at $M_p\sim 1.0
\ME$ and $M_p\sim 1.8 \ME$ in the middle and rightmost panels of the
standoff radius and polar cap area contours.  Planets to the right of
these boundaries still have a completely liquid core and therefore
produce weaker PMFs (lower standoff radius and larger polar cap
areas).  On the other hand, the inner core in planets to the left of
the these boundaries have already started to grow and therefore their
PMFs are stronger.
%REVISED-MAN

Unlocked planets (figure \ref{fig:Bmp-Rs-Apc-Notid}) are better
protected than slowly rotating tidally locked planets by developing
extended magnetospheres $R_S\gs 4 \ R_p$ and lower polar cap areas
$A_\rom{pc}\ls 10\%$.  It is interesting to notice that in both cases
and at times $t\sim 1$ Gyr a smaller planetary mass implies a lower
level of magnetic protection (lower standoff distances and larger
polar caps).  This result seems to contradict the idea that low-mass
planets ($M_p\ls 2$) are better suited to develop intense and
protective PMFs \citep{Gaidos10,Tachinami11,Zuluaga12}.  To explain
this contradiction one should take into account that magnetic
protection as defined in this work depends on dipole moment instead of
surface magnetic field strength.  Since dipole moment scales-up as
${\cal M}\sim B_\rom{dip} R_p^3$ more massive planets will have a better chance to
have large and protective dipole moments.
%REVISED-MAN

It is interesting to compare the predicted values of the maximum
dipole moment calculated here with the values roughly estimated in
previous attempts \citep{Griebmeier05,Khodachenko07,LopezMorales11}.
On one hand \citealt{Khodachenko07} estimate dipole moments for
tidally locked planets in the range 0.022-0.15 ${\cal M}_\oplus$.
These values have been systematically used in the literature to study
different aspects of planetary magnetic protection (see
e.g. \citet{Lammer10} and references therein).  For the same type of
planets our model predicts maximum dipole moments almost one order of
magnitude larger (0.15-0.60 ${\cal M}_\oplus$) with the largest
differences found for the most massive planets ($M\gs 4\ME$).  These
differences arise from the fact that none of the scaling-laws used by
\citealt{Khodachenko07} depend on the convective power.  In our
results the dependency on power explains the differences between
massive planets and ligther planets especially at early times.  On the
other hand \citealt{LopezMorales11} estimate magnetic dipolar moments
of tidally locked super-Earths in the range 0.1-1.0 ${\cal M}_\oplus$.
These values are compatible with our results.  In their, work
\citealt{LopezMorales11} use the same power-based scaling laws we
applied here but assume a rather simple interior model and a static
thermal model where the convective power is set such that maximizes
the efficiency with which the convective energy is converted into the
magnetic field.
%REVISED-FIN

\begin{figure}
  \centering
  \includegraphics[width=0.6
  \textwidth]{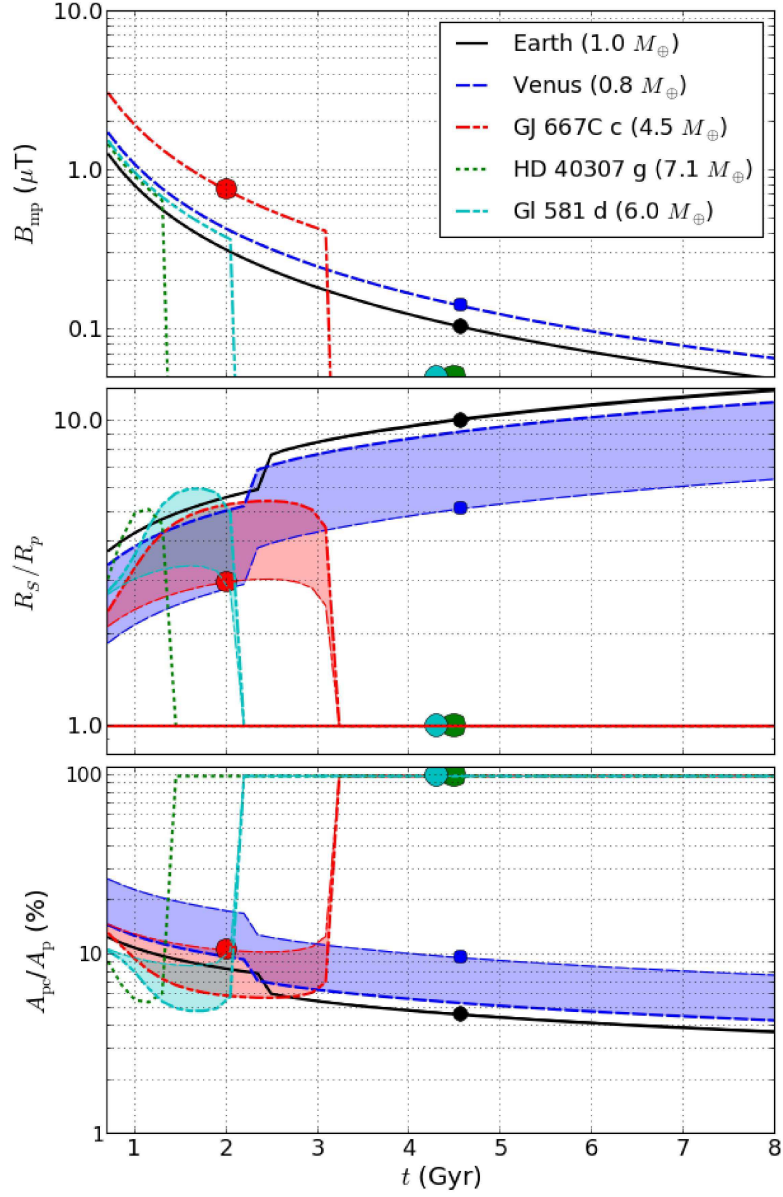}
  \caption{Evolution of magnetosphere properties for the already
    discovered habitable SEs, the Earth and an ``hydrated'' version of
    Venus (low viscosity mantle and mobile lid).  Shaded regions are
    bounded by the properties calculated at a minimum period of
    rotation of $P\approx 1$ day (upper and lower bounds in standoff
    radius and polar cap area curves respectively) and the maximum
    period of rotation $P\approx P_o$ corresponding to a perfect match
    between the rotation and orbital periods (tidal
    locking). Magnetopause fields do not depend on the rotation period
    of the planet.  Filled circles are the predicted present day
    magnetosphere properties computed according to the properties
    summarized in table \ref{tab:SEs}.\vspace{0.5cm}}
  \label{fig:MagnetosphereSEs}
  %REVISED-MAN
\end{figure}

A more detailed account of the evolution of magnetosphere properties
for the already discovered habitable planets is presented in figure
\ref{fig:MagnetosphereSEs}.  In all cases we have assumed that all
planets have compositions similar to Earth (RTEM).  Although almost
all planets are tidally locked, we have also computed the magnetic
properties for a primordial period of rotation $P=1$ day.
%REVISED-MAN

The case of the ``hydrated'' Venus is particularly interesting in
order to analyse the rest of planets.  The dynamo of the actual Venus
probably shut down at $t=3$ Gyr as a consequence of the drying of the
mantle \citep{Christensen09b}.  A massive loss of water induced by a
runaway greenhouse and insufficient early magnetic protection played a
central role in the extinction of the early Venusian PMF.  The
evolution of the PMF in the potentially habitable planets GL 581d,
GJ 667Cc and HD 40307g could have a similar fate.  Their masses are much
larger and therefore their atmospheres are protected by stronger
gravitational fields.
%REVISED-MAN

For planet HD 40307g our reference thermal evolution model predicts a
late shut down of the dynamo $t_{\rm dyn}\sim 4$ Gyr.  According to
our reference model the planet is presently devoid of a dynamo
generated magnetic field.  However, being around a K star ($\Ms\sim
0.7$) the stellar wind and XUV radiation have probably decreased
enough to not represent at present times a real threat for its
atmosphere.
%REVISED-MAN

GL 581d and GJ 667Cc are located in the HZ of dM stars where the stellar
wind pressure and XUV radiation, even at times as late as 4 Gyr, are
intense enough to erode their atmospheres or to make them lose their
water content.  The RTEM predicts that for an Earth-composition GL 581d
at present times had already lost its dynamo and has been exposed for
almost 2.5 Gyr to the harmful effects of the stellar wind and CR.  The
planet however is the most massive of the three planets and probably
has a thick atmosphere able to withstand the continuous agression of
its host star.

Given the estimated age of the GJ 667C system ($t\approx 2$ Gyr), the
RTEM predicts that the planet still has a dynamo (red circle in figure
\ref{fig:MagnetosphereSEs}).  Magnetosphere properties are very close
to that of our ``hydrated'' Venus, 4 Gyr ago.  However its mass is
lower than that of GL 581d and it is located at the inner edge of the
HZ where the exposition to the XUV radiation from its host star (a
young M1 star) could have been enough to induce massive loss of
atmospheric gases including water.  We will come back on these issue
in section \ref{subsec:MassLoss} when we will show how to estimate the
atmospheric mass-loss rate for this particular planet.
%REVISED-MAN

%%%%%%%%%%%%%%%%%%%%%%%%%%%%%%%%%%%%%%%%%%%%%%%%%%%%%%%%%%%%%%%%%%%%%%%%%
\subsection{Toward an estimation of the atmospheric mass-loss}
\label{subsec:MassLoss}
%%%%%%%%%%%%%%%%%%%%%%%%%%%%%%%%%%%%%%%%%%%%%%%%%%%%%%%%%%%%%%%%%%%%%%%%%

Combining the model of magnetosphere evolution developed here with
models of thermal and non-thermal atmospheric escape it would be
possible to estimate the mass-loss rate from atmopsheres of magnetised
and unmagnetised potentially habitable planets.  This is a fundamental
goal to be pursued in the near future if we want to assess the actual
habitability of present and future discovered terrestrial planets in
the HZ of their host stars.  The complex interaction between an
inflated atmosphere and its protective magnetosphere and large
uncertainties in the surface fluxes of atmospheric gasses that
compensate the loss of volatiles induced by the action of the stellar
wind, render this goal hard to achieve in the present.  Despite these
limitations we can still make order of magnitude estimations based on
our own results and the mass-loss rate computed for example in the
recent works by \citealt{Tian08}, \citealt{Tian09} and
\citealt{Lammer12}.
%REVISED-MAN

Atmospheric thermal mass-loss induced by the exposition to X-rays and
EUV stellar radiation (XUV) have been estimated for the case of
Earth-like N$_2$ rich atmospheres \citep{Watson81,Kulikov06,Tian08}
and dry Venus-like CO$_2$ rich atmospheres \citep{Tian09,Lammer12}.
One critical property of an inflated atmosphere is essential to
evaluate the exposition of such atmospheres to further non-thermal
processes: the radius of the exobase $R_\rom{exo}$.  $R_\rom{exo}$ is
defined as the distance where the mean-free path of atmospheric
particles could be comparable to the size of the planet.  When the
radius of the exobase is comparable or larger than the magnetic
standoff distance $R_S$ we will say that the planet is unmagnetised.
Under these conditions the gases escaping from the exosphere will be
picked-up by the stellar wind and lost to the interplanetary space.
On the other hand if the exobase is well inside the magnetosphere
(which is the case of the Earth today) atmopsheric gasses escaping
thermally from the exosphere could stay trapped by the magnetic field
forming a plasmasphere.  Planets under this condition will be
magnetically protected and the mass-loss rate is expected to be much
lower than for unmagnetised planets.
%REVISED-MAN

Using the conservative estimation of the X and EUV luminosities of
main-sequence stars given by \citealt{Garces11} we have estimated the
XUV flux at the top of the atmospheres of GL 581d, GJ 667Cc and HD
40307g during the first critical gigayear of planetary evolution.  The
planet that received the minimum amount of XUV radiation is HD 40307g
with $F_{\rm XUV}=35-10$ PEV (1 PEV = 0.64 erg cm$^{-2}$ s$^{-1}$ is
the Present Earth Value , \citealt{Judge03,Guinan09}).  GL 581d was
exposed in the first gigayear to a flux of $F_{\rm XUV}=150-250$ PEV while
GJ 667Cc received the maximum amount of XUV radiation among them,
$F_{\rm XUV}=450-800$ PEV.
%REVISED-MAN

Using the recent results by \citealt{Tian09} that computed the
exosphere properties of massive super-Earths, i.e. $M_p\geqslant
6\ME$, subject to different XUV fluxes, we can estimate the exosphere
radius and mass-loss rate for our three habitable super-Earths.
Actually, since the \citet{Tian09} results are only available for
planets with a minimum mass of $M_p=6\ME$, a qualitative extrapolation
of the results for 10, 7 and 6 $\ME$ (see figure 4 and 6 in its
paper), shows that exobase radius and mass-loss rates are larger for
less massive planets.  This is particularly useful at trying to apply
the Tian's results to GJ 667Cc $M_p\approx 4.5\ME$ and other less
massive potentially habitable planets.  In these cases we will use the
results by \citet{Tian09} to calculate a lower bound of the exobase
radius and mass-loss rates.
%REVISED-MAN

%% Using the estimated XUV flux for HD 85512b ($M_p\approx 7 \ME$) we
%% predict a minimum exosphere radius at times $t\sim 1$ Gyr between 1.4
%% and 2.0 $R_p$.  Our magnetic protection model for this planet predicts
%% magnetic standoff distances $R_S>2 R_p$ (see figure
%% \ref{fig:MagnetosphereSEs}).  Being the lightest planet the actual
%% exobase radius should be much larger than the minimum computed from
%% the \citet{Tian09} results.  The proximity between the minimum exobase
%% radius and the magnetic standoff distance does not allow us to
%% properly judge the level of magnetic protection of the planet with the
%% information presently available.  Further theoretical investigations
%% will be required to assess the actual magnetic protection of this
%% planet.
%REVISED-MAN

Using the estimated XUV flux for HD 40307g ($M_p\approx 7 \ME$) and
assuming an initial CO$_2$ rich atmosphere, Tian's results predict
that the exosphere of the planet and hence its mass-loss rate was low
enough to avoid a significant early erosion of its atmosphere (see
figures 4 and 6 in his paper).  This is true at least during the first
1-2 Gyr during which our magnetic protection model predict the planet
was enshrouded by a protective magnetosphere (see figure
\ref{fig:MagnetosphereSEs}).  After dynamo shut down the atmosphere of HD 40307g
has been exposed to the direct action of the stellar wind.  Assuming a
stellar age of 4.5 Gyr \citep{Tuomi2012} this effect has been eroding
the atmosphere for 3-4 Gyrs.  During this unmagnetised phase the
atmospheric mass-loss rate can be simply estimated as $\dot M\approx
\alpha m n v_\rom{eff}$ \citep{Zendejas10} where $\alpha$ is the
so-called entrainment efficiency and $m$, $n$ and $v_\rom{eff}$ are
the mass, number density and effective velocity of the stellar wind as
measured at planetary distance (see eq. (\ref{eq:Psw})).  Using a
entrainment efficiency $\alpha\sim 0.3$ (which is appropriate for
example to describe the mass-loss of the Venus atmosphere) we obtain
that the total mass-loss during the unmagnetised phase is less than
1\% of a conservative estimate of the total volatile content of the
planet \citep{Tian09}.  Although our model provides only upper limits
to magnetic protection and the planet could have for example a lighter
Nitrogen-rich atmosphere which is more prone to XUV induced
mass-losses \citep{Watson81,Kulikov06,Tian08}, this preliminary
estimation suggests that HD 40307g probably still preserve a dense
enough atmosphere able to sustain surface liquid water and hence to be
actually habitable.
%REVISED MAN

The case of GL 581d ($M_p\approx 6 \ME$) is quite different.  Assuming
that our estimations of the XUV flux are right, the exosphere radius
predicted by \citealt{Tian09} should be close to the actual one.  In
this case at times $t\sim 1$ Gyr, $R_{\rm exo}=1.8-2.3\;R_p$.  However
our reference magnetosphere model predicts for this planet magnetic
standoff distances $R_S>2.7$ at all times.  We conclude that using our
estimations GL 581d could have been protected by its intrinsic
magnetic field during the critical early phases of planetary evolution
and probably has preserved the critical volatiles in its atmosphere.
Still the uncertainties in the exosphere model or in the atmopsheric
composition and of course in the magnetic model developed here should
lead to a different conclusion and further theoretical and probably
observational investigation is required.
%REVISED-MAN

The most interesting case is that of GJ 667Cc ($M_p\approx 4.5 \ME$).
The minimum exosphere radius predicted for this planet at $t\sim 1$
Gyr lies between $3.0-4.5\;R_p$ while according to our magnetosphere
model, the magnetic standoff distance is $R_S<3\;R_p$ up to 2 Gyr.
Since the exosphere radius should actually be larger than that
predicted with the \citet{Tian09} model and our magnetic model is
actually optimistic, the chances that this planet was unprotected by
its magnetic field in the critical first gigayear are high.  But exposition
does not necessarily mean a complete obliteration of the atmosphere
(see for example the case of Venus).  To evaluate the level of thermal
and non-thermal obliteration of the atmosphere we need to estimate the
actual mass-loss rate.  At the XUV fluxes estimated at the top of the
atmosphere of this planet during the first gigayear, the minimum thermal
mass-loss rate of Carbon atoms from a CO$_2$ rich atmosphere will be
larger than $2-4\times 10^{10}$ atoms cm$^{-2}$ s$^{-1}$ (see figure 6
in \citealt{Tian09}).  We should recall that this is actually the
value for a $6\ME$ super-Earth.  For the actual mass of the planet,
$4.5\ME$, the mass-loss rate could be even larger.  Moreover if as
predicted here the exosphere is exposed directly to the stellar wind,
non-thermal processes can contribute to a larger increase in the
mass-loss from the planetary atmosphere.  At the minimum mass-loss
rate the exposed GJ 667Cc atmosphere could have lost more than $\sim
10^{46}$ atoms of Carbon in just $\sim 100$ Myr and in the first gigayear
the amount of carbon thermally lost to space could rise to $\sim
10^{47}$ atoms.  If we scale-up linearly with planetary mass the total
inventory of CO$_2$ in the atmosphere, crust and mantle of Venus,
which is $2-3 \times 10^{46}$ molecules (see \citealt{Tian09} and
references therein), a $4.5\ME$ planet will have a total budget of
$\sim 10^{47}$ CO$_2$ molecules.  In summary at the minimum mass-loss
rate and assuming a relatively rapid degassing of the planet, Gj 667C
c could have lost its total inventory of Carbon to interplanetary
space in the first couple of gigayears.  Even assuming that large amounts
of CO$_2$ are still trapped in the mantle and crust of the planet, its
atmosphere should be being rapidly obliterated by the stellar wind.
We speculate that GJ 667Cc is a sort of ``Venus-like'' planet.
Regardless the fact the planet is well inside the radiative habitable
zone it has lost its capacity to support life via a massive stellar-wind
induced loss of volatiles.
%REVISED-MAN

%%%%%%%%%%%%%%%%%%%%%%%%%%%%%%%%%%%%%%%%%%%%%%%%%%%%%%%%%%%%%%%%%%%%%%%%%
\section{Discussion}
\label{sec:Discussion}
%%%%%%%%%%%%%%%%%%%%%%%%%%%%%%%%%%%%%%%%%%%%%%%%%%%%%%%%%%%%%%%%%%%%%%%%%

Applying simplified thermal evolution model and dynamo scaling laws to
planets whose bulk properties are barely known or even hypothetical it
is challenging and probably raises more questions than it attempts to
respond.  Further observations of the potentially habitable planets
should be required to precise their physical properties and to
reliably model its interior and thermal evolution.  Moreover continued
observational efforts to look for direct evidence or proxies of
planetary magnetospheres and any other signatures of magnetic
protection, though challenging, should also be attempted.  Here we
want to discuss the assumptions on which our global model relies, its
uncertainties as measured by the sensitivity of the model to changes
in key physical parameters and the missing pieces of information and
present observational limitations to confirm or improve this and other
models of planetary magnetic protection.
%REVISED-FIN

%......................................................................
\subsection{Model assumptions}
\label{subsec:Assumptions}
%......................................................................

The strength of a physical model depends on the hypothesis and
assumptions on which it relies.  Apart from numerous albeit very
common assumptions, the magnetic protection model presented here
depends on three major assumptions we discuss in the following
paragraphs.
%REVISED-NEW

We have assumed that terrestrial planets always develop an initially
metallic liquid core, irrespective of their composition and early
formation history.  This is not necessarily true.  The formation of a
metallic liquid core would depend on very complex processes and other
barely known physical factors.  It has been shown for example that
under extreme water oxidation of iron the formation of a metallic core
will be avoided \citep{Elkins08}.  In this case silicate coreless
planets will be formed.  On the other hand even if a planet is well
differentiated the core could be solid from the beginning (see
e.g. VAL06). However, it should be emphasized here that
our model provides only the best-case scenario of magnetic protection.
Therefore, if under the assumption of having a liquid metallic core, a
planet is found to be lacking enough magnetic protection, the case
when the planet is not well-differentiated or when it never develops a
liquid core will be even worse.  In these cases the conclusions drawn
from our model will be unchanged.
%REVISED-NEW

The calculation of key magnetosphere properties relies on very
simplistic assumptions about the complex physics behind the
interaction between planetary and interplanetary magnetic fields and
stellar wind.  In particular, standoff distances calculated with
eq. (\ref{eq:Rs}) assume a negligible plasma pressure inside the
magnetosphere.  This condition could be violated in planets with very
inflated atmospheres and/or at close-in orbits.  Under these extreme
conditions the magnetic standoff-distance given by eq. (\ref{eq:Rs})
could be a poor underestimation of the actual magnetopause distance.
However, the weak dependence of standoff distances and polar cap areas
of the stellar wind dynamic pressure, offers some idea as to the
actual role that magnetospheric plasma pressure may pay in determining
the size of the magnetosphere. Adding a plasma pressure term
$P_\rom{pl}$ to the magnetic pressure $P_\rom{mp}$ in the left-hand
side of eq. (\ref{eq:BmpPsw}), is equivalent to substracting it from
the stellar wind dynamic pressure $P_\rom{sw}$ in the right-hand
side.  An effective stellar wind pressure $P'_\rom{sw}=P_\rom{sw}
(1-P_\rom{pl}/P_\rom{sw})$ would replace the stellar wind term in the
standoff distance definition (eq. (\ref{eq:Rs})).  As a result a
plasma-pressure correcting factor $(1-P_\rom{pl}/P_\rom{sw})^{-1/6}$
will modify our estimated purely-magnetic standoff distance.  Even in
a case where the plasma was able to exert a pressure 50\% of that of
the stellar wind, the standoff distance will be increased by only a
10\%.  On the other hand in order to have a standoff distance one
order of magnitude larger than that estimated using eq. (\ref{eq:Rs})
the plasma pressure inside the magnetosphere should amount for
99.999\% of the total pressure.  This is precisely what an
unmagnetised planet would look like.  In summary including more
realistic condition into the definition of the magnetosphere boundary
will not modify too much our results.
%REVISED-NEW

A final but not less important assumption in our model is that of
quiet stellar wind conditions.  We have only taken into account
average or quiet stellar wind conditions.  We have completely
neglected the effects of large but transient conditions such as those
produced during coronal mass ejections (CME).  To model the effect
that a steady flux/influx of CME plasma could have in close-in
planets, we can modify the stellar wind pressure by maintaining the
nominal velocity of the plasma but increasing the number density of
wind particles by a factor of two\footnote{Under typical conditions of
  solar CME, the velocity of the wind is not modified too much but the
  plasma densities is increased up to 5-6 times over the average
  particle density}.  Taking into account that $R_S\sim P_{\rm
  SW}^{-1/6}$ we found that under the harsher conditions the
magnetosphere radius and polar cap areas will be modified only by
10-30\% in respect to the nominal or quiet values computed here.  This
simplified estimation shows that our results seem to be robust in
relation to uncertainties in the stellar wind pressure.  However given
the complexity of the interaction between the magnetosphere and the
stellar wind under active phases, a further examination of this case
is required and is left to future research.
%REVISED-MAN

%......................................................................
\subsection{Sensitivity analysis}
\label{subsec:SensitivityAnalysis}
%......................................................................

In order to study the effect that uncertainties in several critical
thermal evolution parameters have in the prediction of the overall
magnetic protection of potentially habitable TPs, we performed a
sensitivity analysis of our model.  For this purpose we independently
varied the value of 6 carefully chosen parameters of the model (see
below) and compared the predicted dipole moment, the time of inner
core formation and the dynamo lifetime with the same values obtained
using the RTEM.
%REVISED-FIN

\begin{figure}
  \centering
   \includegraphics[width=1.0
   \textwidth]{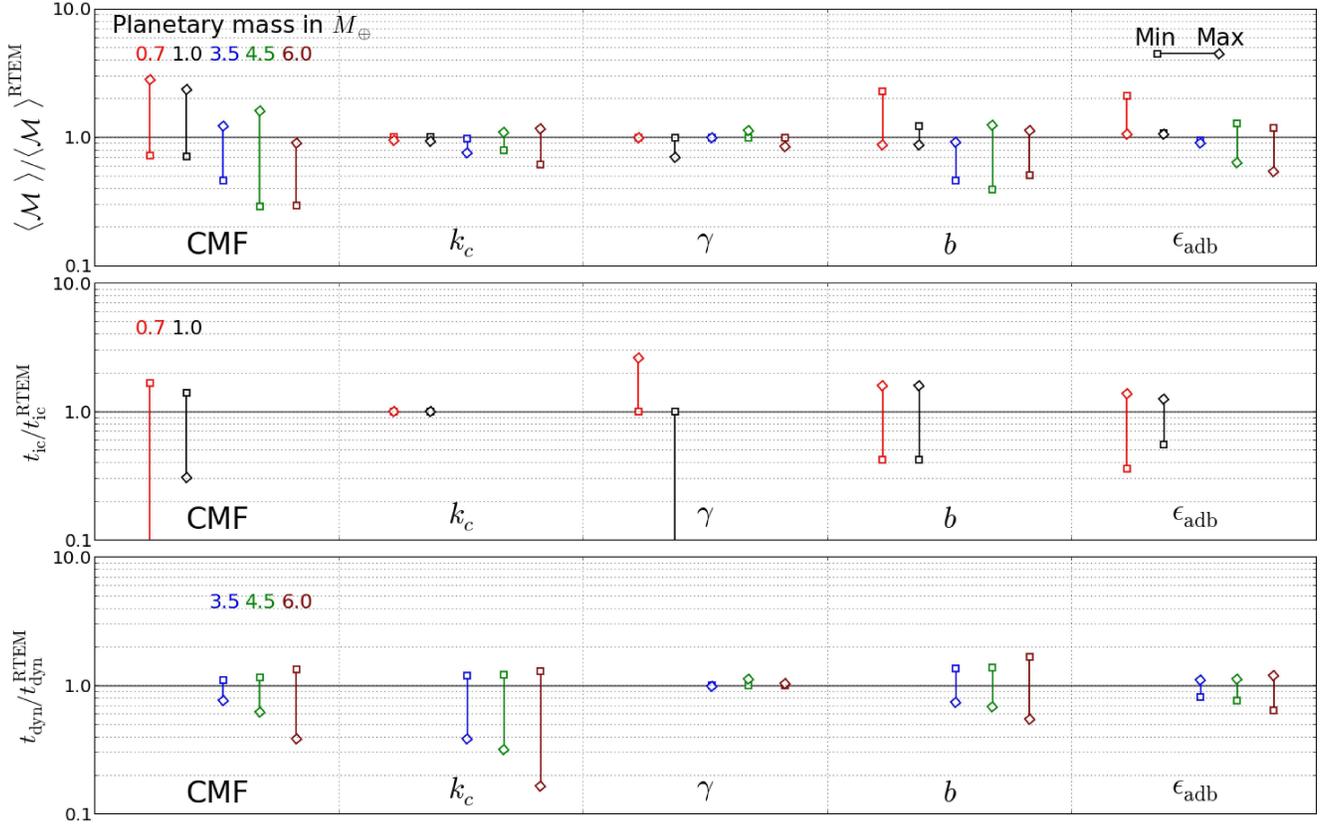}
   \caption{Sensitivity analysis of our reference thermal evolution
     model (RTEM).  Squares and diamonds indicate the relative value
     of three critical magnetic properties, $\langle{\cal M}_{\rm
       dip}\,\rangle$ (the average of the dipole moment between 0.7
     and 2 Gyr), $t_{\rm ic}$ (time of inner core formation) and
     $t_{\rm dyn}$ (dynamo lifetime), as calculated by the thermal
     evolution model.  For the analysis 5 different key thermal
     evolution parameters were independently changed with respect to
     the reference value in the RTEM: the core mass fraction (CMF),
     the thermal conductivity of the core ($k_c$), the Gr\"uneisen
     parameter at core conditions ($\gamma_{0{\rm c}}$), the high
     pressure viscosity rate coefficient ($b$) and the adiabatic
     factor ($\epsilon_{adb}$).  The results obtained when the minimum
     value of the parameters were used are indicated with squares.
     Conversely, the results obtained with the maximum value of each
     parameter are indicated with diamonds.
    \vspace{0.5cm}}
  \label{fig:SensitivityAnalysis}
  %REVISED-MAN
\end{figure}

We performed these comparisons for planets with five different masses:
0.7, 1.0, 3.5, 4.5 and 6.0 $\ME$ (see figure
\ref{fig:SensitivityAnalysis}).  These masses correspond approximately
with those of the already discovered habitable planets (see table
\ref{tab:SEs}) including a hydrated Venus and present Earth.  In all
cases we assume for simplicity a primordial period of rotation of
$P=1$ day.
%REVISED-MAN

Since the dipole moment is an evolving quantity we have plotted in
figure \ref{fig:SensitivityAnalysis} the average value of this
quantity as calculated in the interval 0.7-2.0 Gyr.  For times earlier
than 0.7 Gyr the stellar wind pressure is uncertain and the magnetic
protection cannot be estimated (as discussed in section
\ref{subsec:StellarWind} observations suggest lower stellar wind
pressures at times earlier than this).  For times larger than 2.0 Gyr
the flux of XUV radiation and the stellar wind pressure has decreased
below the initial high levels.  Although an average of the dipole
moment is not phenomenologically relevant, it could be used as a proxy
of the overall magnetic shielding of the planet during the harsh early
phases of stellar and planetary evolution.
%REVISED-MAN

After studying the full set of physical parameters involved in our
interior structure and thermal evolution models (see table
\ref{tab:ThermalModelParameters}) we identified 6 critical parameters
whose values could have noticeable effects on the results or are
subject to large uncertainties.  We performed an analysis of the
sensitivity that the model have to the variation of the following
physical parameters:
%REVISED-MAN

\begin{enumerate}

\item {\bf The core mass fraction, CMF}.  This is the fraction of the
  planetary mass represented by the metallic core.  This parameter is
  determined by the Fe/Si ratio of the planet that it is fixed at
  planetary formation or could be altered by exogenous processes
  (e.g. late large planetary impacts).  Our reference model uses the
  Earth's value CMF = 0.325, i.e. assumes that all planets are
  dominated by a sillicate-rich mantle.  As a comparison Mars has a
  CMF = 0.23 and the value for Mercury is CMF = 0.65 (it should be
  recalled that Mercury could have lost a significant fraction of its
  mantle sillicates increasing the total iron fraction, probably after
  an early large impact).  The CMF determines the size of the core and
  hence the thermal properties of the convective shell where the
  magnetic field is generated.  For our sensitivity analysis we have
  taken two extreme values of this parameter, CMF = 0.23 (a mars-like
  core) and CMF = 0.43 (an iron-rich core).  Planets with larger cores
  have low pressure olivine mantles and our rheological model becomes
  unreliable.
  %REVISED-MAN

\item {\bf The initial temperature contrast at the CMB, $\Delta T_{\rm
    CMB}=\epsilon_{\rm adb}\Delta T_{\rm adb}$} (see
  eq. (\ref{eq:InitialCoreTemperature})).  This is one of the most
  uncertain properties in thermal evolution models.  The initial core
  temperatures would be determined by random processes involved in the
  assembly and differentiation of the planet.  It could vary widely
  from planet to planet.  In order to fit the thermal history of the
  Earth (time of inner core formation, present size of the inner core
  and magnetic field strength) we set $\epsilon_{\rm adb}=0.7$ and
  applied the same value to all planetary masses.  In our sensitivty
  analysis we varied this parameter between two extreme values of 0.6
  and 0.8.  Though we are not sure that this interval is
  representative of planets with very different masses and formation
  histories, our analysis provides at least the magnitude and sign of
  the effect that this parameter has in the dynamo properties
  predicted by our thermal evolution model.
  %REVISED-MAN

\item {\bf High pressure viscosity rate coefficient, $b$} (see
  eq. (\ref{eq:ViscosityModelLower})).  Rheological properties of
  sillicates inside the mantle are among the most uncertain aspects of
  thermal evolution models.  They critically determine, among other
  key quantities, the amount of heat that the core and mantle could
  transport through their respective boundary layers (see
  eqs. (\ref{eq:Qc}), (\ref{eq:MantleHeat}) and
  (\ref{eq:MantleHeat_Stagnant})).  We found that the viscosity of the
  lower mantle (perovskite) is the most important source of
  uncertainties in our thermal evolution model.  The formula used to
  compute viscosity at that layer (see
  eq. (\ref{eq:ViscosityModelUpper})) strongly depends on temperature
  and pressure and the parameter controlling this dependence is the
  ``rate coefficient'' $b$.  In the RTEM we used the value $b=12.3301$
  to reconstruct the thermal properties of the Earth.  In this value
  all the figures are significative reflecting the strong sensitivity
  of the model to this parameter.  To study the impact of $b$ in the
  model results, we varied it in the interval $10-14$.
  %REVISED-MAN 

\item {\bf Iron thermal conductivity, $k_c$}.  This parameter controls
  the amount of heat coming out from the core.  In the RTEM we used a
  value $k_c$=40 Wm$^{-1}$K$^{-1}$ that fits the thermal evolution
  history and present magnetic field of the Earth (see table
  \ref{tab:ThermalModelParameters}).  Although recent first-principles
  analysis suggest that values as large as 150-250 Wm$^{-1}$K$^{-1}$
  could be common at Earth's core conditions \citep{Pozzo12} we
  conform here to the standard values of this parameter.  Further
  investigations to explore values as large as that found by
  \citealt{Pozzo12} should be attempted.  For our sensitivity analysis
  we varied $k_c$ between 35 and 70 Wm$^{-1}$K$^{-1}$, two values
  which are inside the typical uncertainty assumed for this property.
  %REVISED-MAN

\item {\bf Gr\"uneisen parameter for iron, $\gamma_{0c}$}.  This is one
  of the most critical parameters of the equation of state specially
  at core conditions.  It affects strongly the mechanical structure,
  temperature profile and phase of iron in the metallic core (for a
  detailed discussion on the sensitivity of interior structure models
  to this parameter see e.g. VAL06).  In the RTEM we used
  the reference value $\gamma_{0c}$=1.36 that fits the thermal
  evolution history and present magnetic field of the Earth (see table
  \ref{tab:ThermalModelParameters}).  Assuming different kind of core
  alloys a relatively large range of values of this parameter has been
  used in literature (see VAL06 and references there in).
  Gr\"uneisen parameter values have been found in the range of
  1.36-2.338.  Since our RTEM value is at the lower end of this range
  for our sensitivity analysis we have recalculated the model for a
  larger value of 2.06.

%% \item {\bf Critical Rayleigh number at the CMB, $Ra_{c}$}.  This
%%   parameter controls the amount of heat transported through the CMB
%%   and delivered to the mantle (see eq. (\ref{eq:Rac})).  A typical value
%%   $Ra_{c}=1000$ is frequently assumed and was used here in the RTEM.
%%   For our sensitivity analysis we have recalculated the thermal
%%   evolution when the parameter is between 900 and 1100, a typical
%%   interval assumed in the literature.
%%   %REVISED-MAN

\end{enumerate}

Other uncertain parameters such as the critical Rayleigh number at the
CMB, $Ra_{c}$, that is also varied to study the sensitivity of thermal
evolution models \citep{Gaidos11}, were also studied.  We did not
found significant sensitivity of our model to variation of those
parameters.

We depict in figure \ref{fig:SensitivityAnalysis} the relative
variation in the aforementioned magnetosphere and dynamo properties
when each of the previously described parameters were varied
independently.
%REVISED-FIN

We have found that planetary composition (CMF), mantle viscosity are
responsible for the largest uncertainties in the predicted magnetic
properties of the planet. Planets with small metallic cores have on
average low magnetic dipole moments (squares in the first column of
the upper panel).  This is mainly due to a geometrical effect.  The
total heat produced by the core and hence the magnetic field strength
at core surface is of the same order for Fe-poor and Fe-rich planets.
However a small core means also a lower magnetic dipole moment,
i.e. ${\cal M}\sim R_c^3$.  Planets with lower content of iron also
have small and hot cores and therefore the solid inner core formation
and the shutting down of the dynamo are slightly delayed (squares in
the middle and lowest panel of figure \ref{fig:SensitivityAnalysis}).
%REVISED-MAN

Viscosity dependence on pressure and temperature, as quantified by the
parameter $b$, has the opposite effect on planetary magnetic
properties than CMF at least for earth-like planets.  A low viscosity
lower mantle will favour the extraction of heat from the metallic core
increasing the available convective energy for dynamo action.  On the
other hand a viscuous lower mantle will delay the formation of a solid
inner core and extend the lifetime of the dynamo (middle and lowest
panel in figure \ref{fig:SensitivityAnalysis}).
%REVISED-MAN

The effect of the Gr\"uneisen parameter at core conditions are
negligible, at least on what respect to the magnetic field strength
and lifetime (upper and lower panels) which are the most critical
properties affecting planetary magnetic protection.  Only the time of
inner-core formation is strongly affected by changes in this
parameter.  In planets smaller than Earth, inner-core solidification
can be delayed up to three times the reference value.  On the other
with a larger Gr\"uneisen parameter hand, earth-like planets could get
a solid inner-core very early in their thermal histories even almost
since the beginning.  This is the result of the interplay between the
resulting evolution of the thermal profile and the solidus.
%REVISED-MAN

The effect of the initial temperature contrast across the CMB,
quantified by the parameter $\epsilon_{\rm adb}$, goes in the same
direction as viscosity.  The reasons for this behavior are however far
more complex.  A larger initial temperature contrast across the CMB
also implies a larger initial temperature at the core center.
Although a hotter core also produce a larger amount of available
convective energy, the time required for iron to reach the
solidification temperature is also larger.  The dynamo of planets with
$M_p<2 \ME$ and hot cores (large $\epsilon_{\rm adb}$) is weaker than
that of more massive planets during the critical first couple of gigayears
where the average is calculated.  Planets with a colder core develops
a solid inner core almost from the beginning and the release of latent
and gravitational energy feeds a stronger dynamo.
%REVISED-MAN

The case of more massive planets, $M_p>2 \ME$, where the condition for
an inner core formation is never reached during the dynamo lifetime,
is different.  In this case planets with hot cores (large lower mantle
viscosities or high temperature contrasts along the CMB) produce large
amounts of available convective energy.  A larger convective power
will produce a larger value of the local Rossby number.  Thus massive
planets with hot cores also have multipolar dynamos and hence lower
dipole magnetic moments and a reduced magnetic protection.
%REVISED-MAN

Thermal conductivity affect less the results of the thermal evolution
model.  Besides the case of massive planets where differences in the
order of 10-30\% in the magnetic properties are observed when $k_c$
varies between its extremes, the magnetic properties calculated with
our reference thermal evolution model seem very robust against
variations in these two properties.  However, it should be mentioned
that this result applies only when a standard value of $k_c$ is
assumed. Further investigations to explore the recent findings
\citep{Pozzo12} Concerning the possibility that $k_c$ could be larger
by factors as large as 2-3 should be attempted.
%REVISED-MAN

In summary, despite the existence of a natural sensitivity of our
simplified thermal evolution model to uncertainties on their free
parameters, the results presented in this paper seem to be correct at
least in the order of magnitude.  Moreover since the standoff distance
and polar cap areas, which are the actual proxies to magnetic
protection, goes as ${\cal M}^{1/3}$, a one order of magnitude
estimation of ${\cal M}$ will give us an estimation of the level of
magnetic protection off by a factor around 2.
%REVISED-MAN

To clarify this point let's consider the case of GL 581d.  If we
assume for example that its iron content is much less than that of the
Earth (as was assumed in the RTEM model) but the rest of critical
thermal properties are essentially the same, the average standoff
distance (polar cap area) at the critical first gigayear will be off by
only $\sim 30\%$ with respect to the prediction depicted in figure
\ref{fig:MagnetosphereSEs}.  More interesting is to notice that
probably the uncertainties due to the unknown period of rotation
(shaded area in figure \ref{fig:MagnetosphereSEs}) seem to be much
larger than those coming from the uncertainties in the thermal
evolution model.
%REVISED-MAN

%% Dynamo scaling laws has been compared and improved using the
%% well-known properties of the solar system planets and low mass stars.
%% However the dependence on rotation of the dipole field intensity
%% obtained through the local Rossby number still expects observational
%% support and other sources of confirmation.
%% %REVISED-MAN

%% The case of the thermal evolution models is similar.  Simple
%% parametric models as that presented in this work have achieved
%% relatively well at predicting the thermal and magnetic properties of
%% the Earth.  They have been also succesful at predicting the global
%% thermal evolution of Mercury, Venus and Mars.  However its
%% applicability to planets with masses larger than the Earth is still
%% uncertain and awaits for future observational support.
%% %REVISED-MAN

%%%%%%%%%%%%%%%%%%%%%%%%%%%%%%%%%%%%%%%%%%%%%%%%%%%%%%%%%%%%%%%%%%%%%%%%%
\subsection{Observational support}
\label{subsec:Observations}
%%%%%%%%%%%%%%%%%%%%%%%%%%%%%%%%%%%%%%%%%%%%%%%%%%%%%%%%%%%%%%%%%%%%%%%%%

Validating or improving thermal evolution models and dynamo scaling
laws for the case of super-Earths nowadays represents a huge
observational challenge.  The available sensitivity of our best
instruments in the Earth and in space are rather insufficent.  New
and/or improved instruments and observational techniques will be
required to detect, catalogue and compare the thermal and magnetic
properties of low mass planets in the medium to far future.  However
the importance that the detection and characterization of the magnetic
properties of future discovered potentially habitable planets in order
to assess their true habitability clearly justify the effort.
%REVISED-MAN

The first goal seems to be the direct or indirect detection of
super-Earth magnetospheres.  Four methods, some of them already used
in our own solar system, could be devised to achieve this goal: 1) the
detection of radio waves coming from synchrotron and cyclotron
radiation produced by plasma trapped in the magnetosphere, 2) the
detection of a bow shock or a tail of ions produced by the interaction
of the planetary atmosphere and magnetosphere with the stellar wind or
the interplanetary plasma, 3) the detection of planetary auroras and
4) spectroscopic observations of a non-equilibrium atmospheric
chemistry induced by a high flux of CR (this is actually a negative
detection of a magnetosphere).
%REVISED-MAN

The first (radioemission) and second methods (bow shock or tail) have
already been studied in detail \citep{Bastian00, Farrell02,
  Griebmeier07b, Lazio09, Vidotto10,Vidotto11}.  Its reliability, at
least for the case of planets with intense magnetic fields or placed
very close to their host star, has been already tested.
%REVISED-MAN

If synchrotron or cyclotron radiation coming from the magnetosphere of
terrestrial planets could be detected, the power and spectra of the
radiation could be used to measure the magnetic field strength.
However, even with the most sensitive instruments, e.g. the Low
Frequency Array, LOFAR or the Long Wavelength Array LWA, the expected
power and spectra are several orders of magnitude below the threshold
of detection.  Powers as large as $10^3$ - $10^5$ times the Jupiter
radioemission and frequencies in the range of tens of MHz are required
for the present detection of synchrotron radio emission in planetary
magnetospheres (see e.g. \citealt{Griebmeier07b}).  The magnetic field
intensities and expected frequencies produced in super-Earths
magnetospheres are several orders of magnitude lower than these
thresholds and probably are far from being detected in the near
future.
%REVISED-MAN

It has been shown recently that measurements of the asymmetry in the
ingress and egress of transiting planets can be used to detect the
presence of a bow shock or a tail of plasma around the planet.  Vidoto
(2010, 2011) used this phenomenon to constrain the magnetic properties
of Wasp-12b.  The formation of a detectable bow-shock depends, among
other factors, on the relative velocity between the planet and the
shocked plasma.  Close-in planets with strong enough magnetic fields
(this is precisely the case of Wasp-12b) can easily develop UV-opaque
bow shocks and allow reliable detection.  However low-mass planets
with relatively weak magnetic fields such as those predicted with our
models, hardly produce a detectable bow-shock.  It has been estimated
that magnetopause fields in the range of several Gauss should be
required to have a detectable signal of a bow shock (A.A. Vidotto,
Private Communication).  Our habitable super-Earths have magnetopause
fields in the order of a few micro Gauss (see figure
\ref{fig:Bmp-Rs-Apc-Tid}).  The case of an ion tail coming from a
weakly magnetised planet has received less attention and probably
could offer better chances for a future indirect detection of the
magnetic environment of low mass planets.
%REVISED-MAN

Finally the detection of far UV (FUV) or X-ray emission from planetary
auroras can also be used as a tool to study directly and indirectly
planetary magnetospheres.  Planetary auroras with intensities as high
as $10^2$ to $10^3$ times larger than that of the Earth are expected
in close-in giant planets subject to the effects of CMEs from its host
star \citep{Cohen11}.  If we estimate that a typical Earth Aurora has
an intensity of 1 kR \citep{Neudegg01} (Being 1 R $\sim 10^{-11}$
photons m$^{-2}$s$^{-1}$srad$^{-1}$) and assuming that 10\% of a
close-in Jupiter-like planet is covered by auroras producing FUV
photons around $130$ nm, the total emitted power from these planets
will be $\sim 10^{13}$ W.  If we assume that this is the present
threshold for exoplanetary aurora detection, even under strong stellar
wind conditions and distances typical of the habitable zone, the total
FUV power produced by auroras in the polar cap of earth-like planets
could be only $10^5$ W which is 8 orders of magnitude less than the
present detection threshold.  Not to mention that the FUV radiation
should be detected against an intense UV background coming from a
probably young and active low mass star.  If we can find ways to
overpass these difficulties, the observation of the FUV and X-ray
emmision and its variability from auroras in potentially habitable
super-Earths could be used as powerful probes of the magnetic
environment around the planet.
%REVISED-MAN

%%%%%%%%%%%%%%%%%%%%%%%%%%%%%%%%%%%%%%%%%%%%%%%%%%%%%%%%%%%%%%%%%%%%%%%%%
\section{Summary and Conclusions}
\label{sec:Conclusions}
%%%%%%%%%%%%%%%%%%%%%%%%%%%%%%%%%%%%%%%%%%%%%%%%%%%%%%%%%%%%%%%%%%%%%%%%%

We studied here the influence that the thermal evolution of
potentially habitable terrestrial planets has in the protection that
an evolving planetary magnetopshere could provide against the
atmospheric erosion caused by the stellar wind.
%REVISED-MAN

We developed a simple parametrized thermal evolution model able to
reproduce the global thermal history and magnetic properties of the
past and present Earth.  We applied this model to predict the thermal
histories of planets with masses in the range of 0.5 to 6.0 $\ME$ and
with chemical compositions similar to Earth.  Using these results and
applying up-to-date dynamo scaling laws we predicted the magnetic
properties of terrestrial planets in the habitable zone as a function
of time, planetary mass and rotation rate.  A simple model of the
evolution and interaction of the stellar wind with the planetary
magnetic field, that has been adapted from previous works, allowed us
to compute the global properties of the magnetosphere in order to
assess the level of magnetic protection that potentially habitable
Earth-like planets could actually have.
%REVISED-MAN

We applied our model to the case of already known potentially
habitable terrestrial planets (GL 581d, HD 40307g and GJ 667Cc), to
the Earth itself and to the case of a hydrated Venus.  In the case of
the Earth our model reproduces fairly well the early and present
thermal and magnetic properties of our planet.  In the case of the
hydrated Venus, the model predicts low values the standoff distance
and large polar cap areas in the frist critical gigayear of planetary and
thermal evolution, which are compatible with the idea that the planet
lacked a strong enough magnetic protection able to avoid a massive
loss of water and volatiles that finally lead to the shut down of its
dynamo $\sim 3$ Gyr ago.
%REVISED-MAN

Compelling results were found in the case of the three already
discovered extra-solar-system potentially habitable planets.  Assuming
an earth-like composition and thermal evolution parameters similar to
those used in the case of the Earth (reference thermal evolution
model, RTEM), our model predicted that the dynamo of GL 581d and HD
40307g have been already shut down.  A younger GJ 667Cc seems to still
have an active dynamo.
%REVISED-MAN

A non-trivial dependence of the magnetic properties on planetary age,
planetary mass and period of rotation has been found in general for
terretrial planets inside the HZ of their host stars.  Thermal
evolution is responsible for the non-trivial relationship among all
these properties.  Contrary to what was found in previous work,
tidally locked planets could develop relatively intense magnetic
fields and extended magnetospheres.  However they also have extended
polar caps and probably multipolar magnetic fields where field lines
open to the interplanetary space and magnetotail regions probably
increasing the non-thermal mass-losses.
%REVISED-MAN

Using recent results for the relationship between the exposition to
XUV radiation, the exobase radius and mass-loss rate from massive
super-Earths, we estimated the level of exposure and mass-losses for
the three already discovered potentially habitable super-Earths.  With
the available information not too much could be said about the
magnetic protection of HD 40307g.  Further theoretical investigations
are required to evaluate this case.  Our model predicts a large enough
magnetosphere able to protect GL 581d against the erosive action of
the stellar wind during the first critical phases of planetary
evolution.  However since our model is still optimistic further
theoretical and probably observational analyses should be performed to
establish on a more solid basis the magnetic protection of this
planet.  Our upper-limit to the standoff-distance and the most
optimistic estimation of exobase radius and mass-loss rate from the
atmosphere of GJ 667Cc, point-out the fact that this planet has
already lost a large fraction of its inventory of volatiles.  All the
evidence compiled in this work make GJ 667Cc a sort of ``Venus
analogue''.  Although further theoretical analyses are required our
best guess is that despite the fact that it is inside the radiative HZ
of its host star the planet is presently uninhabitble.
%REVISED-MAN

Last but not least we tested the robustness of our conclussions by
changing several of the most sensitive input parameters of our thermal
evolution model.  We found that even under the present uncertainties
the predicted properties of planetary magnetopsheres are rather
robust.  We calculated that introducing large variations in the
composition of the planets and the rheological and thermal properties
of their interiors with respect to the reference thermal evolution
model, the critical magnetic properties, such as the standoff radius
and the area of the polar cap, change only by a factor of two.  Results
are also robust against uncertainties in the stellar wind properties
that could be very important in the case of close-in habitable planets
around active and young dM stars.
%REVISED-MAN

The problem of evaluating the magnetic protection of potentially
habitable planets is far from being settled.  Other sources of
intrinsic magnetic fields, thermal evolution and interior structure of
planets with ``exotic'' compositions, improved theoretical models and
new experimental evidence of the behavior of iron at high pressures
and temperatures, improved and validated models of the evolution and
spatial structure of stellar winds and of course more and better
observational data coming from the already discovered habitable
super-Earths and future discovered potentially habitable exoplanets,
will allow us to assess the actual magnetic protection of potentially
habitable environments.
%REVISED-MAN

% %%%%%%%%%%%%%%%%%%%%%%%%%%%%%%%%%%%%%%%%%%%%%%%%%%%%%%%%%%%%%%%%%%%%%%
% \section*{Acknowledgments}
% %%%%%%%%%%%%%%%%%%%%%%%%%%%%%%%%%%%%%%%%%%%%%%%%%%%%%%%%%%%%%%%%%%%%%%

\acknowledgments

We appreciate the useful discussion and comments of Mercedes
Lopez-Morales and other colleagues participating in the {\it Exploring
  Strange New Worlds 2011 Conference} (Arizona, U.S.)  and in the {\it
  VI Taller de Ciencias Planetarias 2012} (Montevideo, Uruguay).
Special thanks to Ignacio Ferrin who with his clever questions and
suggestions originally motivate us to pursue some of the goals of this
work.  We also thank to Lisa Kaltenegger and Jeffrey Linsky for their
insightful comments on preliminary versions of this manuscript.  We
want to give special thanks to all our fellow colleagues abroad that
have provided us with some key literature unobtainable from our
country.  We are grateful to Aaron West and Luke Webb for his careful
revision of the English in the manuscript.  The remnant errors are all
ours.  Anonymous referee contributed significantly not only to the
improvement of the manuscript but to the quality of the research
conclussions and should be also aknowledged. PC is supported by the
Vicerrectoria de Docencia of the University of Antioquia.  This work
has been completed with the financial support of the CODI-UdeA under
the grant IN591CE and by the University of Medellin under the grant
number 530.
%REVISED2

%%%%%%%%%%%%%%%%%%%%%%%%%%%%%%%%%%%%%%%%%%%%%%%%%%%%%%%%%%%%%%%%%%%%%%%%%%%%%%%%%
%BIBLIOGRAPHY
%%%%%%%%%%%%%%%%%%%%%%%%%%%%%%%%%%%%%%%%%%%%%%%%%%%%%%%%%%%%%%%%%%%%%%%%%%%%%%%%%

%%%%%%%%%%%%%%%%%%%%%%%%%%%%%%%%%%%%%%%%%%%%%%%%%%%%%%%%%%%%%%%%%%%%%%%%%%%%%%%%%
%FIGURES
%%%%%%%%%%%%%%%%%%%%%%%%%%%%%%%%%%%%%%%%%%%%%%%%%%%%%%%%%%%%%%%%%%%%%%%%%%%%%%%%%

%FFFFFFFFFFFFFFFFFFFFFFFFFFFFFFFFFFFFFFFFFFFFFFFFFFFFFFFFFFFFFFFFFFFFF
%FIGURE 0
%FFFFFFFFFFFFFFFFFFFFFFFFFFFFFFFFFFFFFFFFFFFFFFFFFFFFFFFFFFFFFFFFFFFFF
%FFFFFFFFFFFFFFFFFFFFFFFFFFFFFFFFFFFFFFFFFFFFFFFFFFFFFFFFFFFFFFFFFFFFF

%FFFFFFFFFFFFFFFFFFFFFFFFFFFFFFFFFFFFFFFFFFFFFFFFFFFFFFFFFFFFFFFFFFFFF
%FIGURE 1
%FFFFFFFFFFFFFFFFFFFFFFFFFFFFFFFFFFFFFFFFFFFFFFFFFFFFFFFFFFFFFFFFFFFFF
%FFFFFFFFFFFFFFFFFFFFFFFFFFFFFFFFFFFFFFFFFFFFFFFFFFFFFFFFFFFFFFFFFFFFF

%FFFFFFFFFFFFFFFFFFFFFFFFFFFFFFFFFFFFFFFFFFFFFFFFFFFFFFFFFFFFFFFFFFFFF
%FIGURE 2
%FFFFFFFFFFFFFFFFFFFFFFFFFFFFFFFFFFFFFFFFFFFFFFFFFFFFFFFFFFFFFFFFFFFFF
%FFFFFFFFFFFFFFFFFFFFFFFFFFFFFFFFFFFFFFFFFFFFFFFFFFFFFFFFFFFFFFFFFFFFF

%FFFFFFFFFFFFFFFFFFFFFFFFFFFFFFFFFFFFFFFFFFFFFFFFFFFFFFFFFFFFFFFFFFFFF
%FIGURE 3
%FFFFFFFFFFFFFFFFFFFFFFFFFFFFFFFFFFFFFFFFFFFFFFFFFFFFFFFFFFFFFFFFFFFFF
%FFFFFFFFFFFFFFFFFFFFFFFFFFFFFFFFFFFFFFFFFFFFFFFFFFFFFFFFFFFFFFFFFFFFF
%%     : $\sim$1 Gyr for
%%     0.5$\ME$, $\sim$ 2 Gyr for 1.0 $\ME$ and $\gs$4.6 Gyr for $M_p\gs$
%%     1.5 $\ME$.\vspace{0.2cm}}

%FFFFFFFFFFFFFFFFFFFFFFFFFFFFFFFFFFFFFFFFFFFFFFFFFFFFFFFFFFFFFFFFFFFFF
%FIGURE 4
%FFFFFFFFFFFFFFFFFFFFFFFFFFFFFFFFFFFFFFFFFFFFFFFFFFFFFFFFFFFFFFFFFFFFF
%FFFFFFFFFFFFFFFFFFFFFFFFFFFFFFFFFFFFFFFFFFFFFFFFFFFFFFFFFFFFFFFFFFFFF

%FFFFFFFFFFFFFFFFFFFFFFFFFFFFFFFFFFFFFFFFFFFFFFFFFFFFFFFFFFFFFFFFFFFFF
%FIGURE 5
%FFFFFFFFFFFFFFFFFFFFFFFFFFFFFFFFFFFFFFFFFFFFFFFFFFFFFFFFFFFFFFFFFFFFF
%FFFFFFFFFFFFFFFFFFFFFFFFFFFFFFFFFFFFFFFFFFFFFFFFFFFFFFFFFFFFFFFFFFFFF

%FFFFFFFFFFFFFFFFFFFFFFFFFFFFFFFFFFFFFFFFFFFFFFFFFFFFFFFFFFFFFFFFFFFFF
%FIGURE 6
%FFFFFFFFFFFFFFFFFFFFFFFFFFFFFFFFFFFFFFFFFFFFFFFFFFFFFFFFFFFFFFFFFFFFF
%FFFFFFFFFFFFFFFFFFFFFFFFFFFFFFFFFFFFFFFFFFFFFFFFFFFFFFFFFFFFFFFFFFFFF
%%   Having a large period of rotation Venus has
%%      properties similar to that expected for tidally locked planets.
%%      The estimation of the magnetospheric properties of an
%%      hypothetical hydrated Venus has also been included.

%FFFFFFFFFFFFFFFFFFFFFFFFFFFFFFFFFFFFFFFFFFFFFFFFFFFFFFFFFFFFFFFFFFFFF
%FIGURE 7
%FFFFFFFFFFFFFFFFFFFFFFFFFFFFFFFFFFFFFFFFFFFFFFFFFFFFFFFFFFFFFFFFFFFFF
%FFFFFFFFFFFFFFFFFFFFFFFFFFFFFFFFFFFFFFFFFFFFFFFFFFFFFFFFFFFFFFFFFFFFF

%FFFFFFFFFFFFFFFFFFFFFFFFFFFFFFFFFFFFFFFFFFFFFFFFFFFFFFFFFFFFFFFFFFFFF
%FIGURE 8
%FFFFFFFFFFFFFFFFFFFFFFFFFFFFFFFFFFFFFFFFFFFFFFFFFFFFFFFFFFFFFFFFFFFFF
%FFFFFFFFFFFFFFFFFFFFFFFFFFFFFFFFFFFFFFFFFFFFFFFFFFFFFFFFFFFFFFFFFFFFF
%%   The
%%     actual curves should lie inside the shaded regions probably closer
%%     to the lower (upper) limits in the case of close-in already
%%     tidally locked planets (including Venus)

%FFFFFFFFFFFFFFFFFFFFFFFFFFFFFFFFFFFFFFFFFFFFFFFFFFFFFFFFFFFFFFFFFFFFF
%FIGURE 9
%FFFFFFFFFFFFFFFFFFFFFFFFFFFFFFFFFFFFFFFFFFFFFFFFFFFFFFFFFFFFFFFFFFFFF
%FFFFFFFFFFFFFFFFFFFFFFFFFFFFFFFFFFFFFFFFFFFFFFFFFFFFFFFFFFFFFFFFFFFFF

\end{document}